\documentclass[a4paper,11pt]{article}
\pdfoutput=1
\usepackage{amssymb,amsmath,amsfonts,makeidx,placeins,pbox,multirow}
\usepackage{graphicx,rotate,subcaption,color,slashed,cite,caption,epstopdf,verbatim}
\usepackage{longtable,tabu}
\usepackage{array}
\usepackage{color}
\usepackage{url}
\usepackage[table]{xcolor}
\usepackage{amsmath}
\usepackage{mathtools}
\usepackage{amssymb}
\usepackage{dsfont}
\usepackage[colorlinks=true,
            linkcolor=blue,
            citecolor=blue,
            urlcolor=blue]{hyperref}
\usepackage[normalem]{ulem}
\usepackage{graphicx}
\usepackage{svg}
\usepackage[justification=centering]{caption}
\usepackage{subcaption}
\usepackage{booktabs}
\usepackage{makecell}  
\newcolumntype{P}[1]{>{\centering\arraybackslash}p{#1}}

\def\lapp{\mathrel{\rlap{\raise.5ex\hbox{$<$}}
                    {\lower.5ex\hbox{$\sim$}}}}
\def\gapp{\mathrel{\rlap{\raise.5ex\hbox{$>$}}
                  {\lower.5ex\hbox{$\sim$}}}}

\usepackage{color}
\long\def\/*#1*/{}
\usepackage{color}
 \usepackage[normalem]{ulem}
\definecolor{darkgreen}{cmyk}{1,0,1,0.4}
\definecolor{darkred}{cmyk}{0,1,1,0.4}
\definecolor{rosso}{cmyk}{0,1,1,0.4}
\definecolor{rossos}{cmyk}{0,1,1,0.55}
\definecolor{rossoc}{cmyk}{0,1,1,0.2}
\definecolor{blu}{cmyk}{1,1,0,0.3}
\definecolor{blus}{cmyk}{1,1,0,0.6}
\definecolor{bluc}{cmyk}{1,1,0,0.1}
\definecolor{verde}{cmyk}{0.92,0,0.59,0.25}
\definecolor{verdec}{cmyk}{0.92,0,0.59,0.15}
\definecolor{verdes}{cmyk}{0.92,0,0.59,0.4}
\definecolor{grigio}{cmyk}{0,0,0,0.07}
\definecolor{rosa}{cmyk}{0,0.1,0.1,0.02}
\definecolor{rosino}{cmyk}{0,0.05,0.05,0.02}
\definecolor{rosas}{cmyk}{0,0.3,0.25,0.05}
\definecolor{celeste}{cmyk}{0.1,0,0,0.02}
\definecolor{giallino}{cmyk}{0,0,0.4,0.02}
\definecolor{rosso}{cmyk}{0,1,1,0.4}
\definecolor{rossos}{cmyk}{0,1,1,0.55}
\definecolor{rossoc}{cmyk}{0,1,1,0.2}
\definecolor{blu}{cmyk}{1,1,0,0.3}
\definecolor{bluc}{cmyk}{1,1,0,0.1}
\definecolor{blucc}{cmyk}{0.7,0.5,0,0}
\definecolor{viola}{cmyk}{0,1,0,0.6}
\definecolor{viola2}{cmyk}{0,1,0.2,0.6}
\definecolor{verde}{cmyk}{0.92,0,0.59,0.25}
\definecolor{verdec}{cmyk}{0.92,0,0.59,0.15}
\definecolor{verdes}{cmyk}{0.92,0,0.59,0.4}
\definecolor{verdino}{cmyk}{0.12,0,0.09,0.05}
\definecolor{giallo}{cmyk}{0,0,1,0}
\definecolor{gialloverde}{cmyk}{0.44,0,0.74,0}

\evensidemargin=\oddsidemargin
\def\bar {\overline}

\def\bea {\begin{eqnarray}}
\def\eea {\end{eqnarray}}

\def\beq{\begin{equation}}
\def\eeq{\end{equation}}
\def\barr{\begin{array}}
\def\earr{\end{array}}

\newcommand{\Tr}{\operatorname{Tr}}

\def\beq{\begin{equation}}
\def\eeq{\end{equation}}

\newcommand{\nc}{\newcommand}

\nc{\hi}{H}
\nc{\hit}{\widetilde{H}}
\nc{\hij }{\mbox{${\hi^\dag i\,\raisebox{2mm}{\boldmath ${}^\leftrightarrow$}\hspace{-4mm} D_\mu\,\hi}$}}
\nc{\hijt}{\mbox{${\hi^\dag i\,\raisebox{2mm}{\boldmath ${}^\leftrightarrow$}\hspace{-4mm} D_\mu^{\,a}\,\hi}$}}

\def\gev{\rm GeV}
\def\tev{\rm TeV}

\def\gev{\,\ensuremath{\mathrm{Ge\kern -0.1em V}}}
\def\tev{\,\ensuremath{\mathrm{Te\kern -0.1em V}}}

\DeclareUnicodeCharacter{2212}{-}

\begin{document}

\begin{center}
{\Large {\bf Learning holographic QCD with unflavored meson spectra}} \\
\vspace*{0.8cm} {\sf Mathew Thomas Arun \footnote{mathewthomas@iisertvm.ac.in}, Ritik Pal\footnote{ritik24@iisertvm.ac.in }} \\
\vspace{10pt} {\small } {\em  School of Physics, Indian Institute of Science Education and Research, Thiruvananthapuram 695551, Kerala, India}
\normalsize
\end{center}
\bigskip
\begin{abstract}
We develop a data-driven neural network framework to reconstruct the five-dimensional background geometry, the dilaton potential, and the chiral-symmetry-breaking scalar potential of holographic QCD from hadron mass spectra.  Framed as an inverse problem, the model is trained using a discretized form of the Schr\"odinger-like equation, which resembles a linear moose in ``deconstructed" 5 dimensions with Dirichlet boundary conditions, in contrast to the AdS/DL with ``emergent" space-time. Using the masses of the unflavored mesons $\rho$, $a_1$, $a_2$, and $f_0$ and their excitations as training data, the model learns confining effective potentials and computes a dilaton profile that satisfies the null energy condition. The network predicts that the dilaton's IR behavior will be much steeper than its quadratic form. Moreover, the symmetry-breaking bulk potential of the scalar field, $V(X) \sim k_1 X^3+k_2 X^4$, was computed, and the parameters $k_1$ and $k_2$ predicted to be $\sim -4$ and $\sim 9$ respectively. The deep-learned parameters, metric, and the dilaton profile were then used to predict the pion mass and its spectrum with good accuracy. A Python code, along with the trained models, is provided to facilitate further studies\footnote{Available at Github, https://github.com/rp-winter/NN-AdS-QCD}.
\end{abstract}

\section{Introduction}


The AdS/CFT correspondence \cite{Maldacena:1997re, Gubser:1998bc, Witten:1998qj} has provided a powerful framework to study strongly coupled gauge theories using classical gravitational theories in higher-dimensional spacetime. One of the most prominent applications of this correspondence is in the study of Quantum Chromodynamics (QCD) through holographic models. The original proposal by Sakai and Sugimoto~\cite{Sakai:2004cn,Sakai:2005yt} with $D8/D4/\bar{D8}$ configuration of the branes to achieve chiral symmetry and its spontaneous breaking, despite its success, could not produce massive pions. This geometry was deformed to $D8/D4/\bar{D8}/D6$~\cite{Kruczenski:2003uq, Hashimoto:2008sr}, to mimic the techni-colour QCD-like sector on the $D6$ brane, which successfully provided explicit chiral symmetry breaking through a bare quark mass term arising from the asymptotic distance between the $D4$ and $D6$ branes. While such techni-colour models, for chiral symmetry breaking, have rich phenomenology and were studied in the context of raising the axion mass~\cite{Holdom:1982ex,Holdom:1985vx}, they remain disfavoured with the discovery of the Higgs.

The bottom-up AdS/QCD framework~\cite{Erlich_2005, Hirn:2005vk,Karch:2006pv,Csaki:2006ji,Falkowski:2006uy,Shock:2006gt,Karch:2010eg}, in contrast to the string-theoretic approach, remains successful in explaining phenomenological signatures, including the mass spectra of mesons and their decay constants. Unlike full string-theoretic duality, parameters and interactions are chosen to reproduce low-energy QCD observables phenomenologically. These models aim to capture the essential features of QCD, such as confinement and chiral symmetry breaking, by constructing a dual gravitational theory in 5-dimensional Anti-de Sitter (AdS) space, related to the 4-dimensional gauge theory via an effective dictionary. Although the exact gravity dual of QCD and the chiral symmetry breaking dynamics remain unknown, phenomenological models successfully reproduce various hadronic properties, such as the meson mass spectrum \cite{Karch:2006pv, Gherghetta_2009, Zhang_2010, PhysRevD.81.014024}, which are identified with the Kaluza-Klein (KK) modes of the bulk fields. The dynamics of these bulk fields are governed by Sturm-Liouville-type differential equations, and their eigenvalues correspond to these masses in the dual theory. These equations can often be transformed into Schr\"odinger-like equations \cite{Karch:2006pv}, allowing the application of quantum mechanical techniques \cite{PhysRevB.101.245139}.

There are several proposals for the functional forms of the background fields (such as the dilaton profile, the warp factor, and the scalar vacuum expectation value), each characterized by a few free parameters, which are then fitted to match the hadronic spectra. For example, the Gherghetta-Kaputsa-Kelley (GKK) model \cite{Gherghetta_2009} takes the standard AdS warp factor $A(z) = -log(z)$, and assumes an ansatz for the bulk scalar vacuum expectation value (VEV) field to reconstruct the dilaton profile. In many improved holographic QCD models, with 5D graviton–dilaton (or graviton–dilaton–scalar) action \cite{PhysRevD.79.075019,PhysRevD.108.106016,PhysRevD.100.106009}, solving the corresponding Einstein-dilaton equations generates a self-consistent warp factor and dilaton profile, instead of an {\it ad hoc} selection.

Although these models can reproduce the low-lying hadronic spectra with reasonable accuracy, they are often limited by the assumptions made about these background field profiles. Neural networks, as universal function approximators \cite{HORNIK1989359}, can approximate these background fields without making any assumptions about their functional forms. There has been an increased interest in the subject with the establishment of a correspondence between the AdS/CFT and Deep Learning (DL)~\cite{Hashimoto:2019bih,Jeong:2025omu}, where the deep layers of a deep Boltzmann machine, from the viewpoint of a 4-dimensional QFT at the visible layer, are seen as an ``emergent" bulk space coordinate. The trained weights serve as the metric for the extra-dimensional space-time. Further, the matching of the DL with holographic QCD~\cite{Hashimoto:2018bnb,Akutagawa_2020,Hashimoto_2021}, leverages this correspondence to model the background geometry by fitting the $\rho$ and $a_2$ mesons using neural networks. This framework was then improved by including a dilaton potential to predict the $\rho$ meson spectrum~\cite{Hashimoto:2021ihd}.

Beyond these, there has been growing interest in the applications of machine learning techniques in Holographic QCD \cite{PhysRevResearch.2.023369, Filev:2025mbt, Ahn:2024jkk}. In particular, machine learning has been employed in the study of glueball spectra \cite{wyr3-2kc5}, heavy-quark potentials \cite{Luo:2024iwf}, heavy quarkonium spectra and their decay constants \cite{zhang2026heavyquarkoniumspectrumdecay}, probing proton structure \cite{kou2026probingprotonstructurephysicsguided}, and using lattice QCD data to learn the QCD phase diagram \cite{Cai:2024eqa}. These developments collectively demonstrate that holographic QCD can be effectively formulated as an inverse problem, with machine learning providing a flexible framework for exploring the space of admissible background configurations.


\begin{table}[!ht]
    \centering
    {\renewcommand{\arraystretch}{1.25}
    \begin{tabular}{|c|c|c|}
    \hline
    \textbf{Meson} & \textbf{State} & \textbf{Experimental Mass (GeV)}  \\
    \hline
    $\rho$  
        & $1$ &  $0.775 \pm 0.001$ \cite{ParticleDataGroup:2024cfk}\\
        & $2$ &  $1.465 \pm 0.025$ \cite{ParticleDataGroup:2024cfk}\\
        & $3$ &  $1.720\pm 0.020$ \cite{ParticleDataGroup:2024cfk}\\
        & $4^\prime$ & $1.909 \pm 0.030$ \cite{BaBar:2007ceh}\\
        & $5^\prime$ & $2.095 \pm 0.004$ \cite{SND:2023gan}\\
    \hline
    $a_1$  
        & 1 &  $1.230 \pm 0.040$ \cite{ParticleDataGroup:2024cfk}\\
        & 2 &  $1.655 \pm 0.016$ \cite{ParticleDataGroup:2024cfk}\\
    \hline
    $a_2$  
        & 1 &  $1.318\pm 0.001$ \cite{ParticleDataGroup:2024cfk}\\
        & 2 &  $1.706 \pm 0.014$ \cite{ParticleDataGroup:2024cfk}\\
    \hline
    $f_0$  
        & $1$ &  $0.478^{+0.029}_{-0.028} $ \cite{PhysRevLett.86.770}  \\
        & $2$ &  $0.990 \pm 0.020$  \cite{ParticleDataGroup:2024cfk}\\
        & $3^\prime$ &  $1.472 \pm 0.012$ \cite{WA76:1991kef}\\
        & $4$ &  $1.522 \pm 0.025$ \cite{ParticleDataGroup:2024cfk}\\
        & $5$ &  $1.733 ^ {+0.008}_{-0.007}$ \cite{ParticleDataGroup:2024cfk}\\
        & $6$ &  $1.982 ^ {+0.054}_{-0.003}$ \cite{ParticleDataGroup:2024cfk}\\
        & $7^\prime$ &  $2.187 \pm 0.014$ \cite{ParticleDataGroup:2024cfk}\\
        & $8^\prime$ &  $2.340 \pm 0.02$  \cite{Sarantsev:2021ein} \\
        & $9^\prime$ &  $2.470 ^{+0.006} _{-0.007}$ \cite{BESIII:2022zel}  \\
    \hline
    $\pi$
        & $1^\prime$ & $0.135 \pm 0.001$ \cite{ParticleDataGroup:2024cfk}\\
        & $2^\prime$ & $1.300 \pm 0.100$ \cite{ParticleDataGroup:2024cfk}\\
        & $3^\prime$ & $1.81 ^{+0.009}_{-0.011}$ \cite{ParticleDataGroup:2024cfk}\\
    \hline
    \end{tabular}
    }
    \caption{Experimental masses (Breit-Wigner) of the mesons. The unprimed numbers refer to states whose masses we have used for training, and the primed numbers refer to the states that we predict using the trained model. }
    \label{tab:inputs}
\end{table}

In this article, we develop a neural network framework to learn the background geometry of the AdS/QCD model, the chiral symmetry breaking potential, dilaton profile, and the effective potentials of the mesons by fitting the mass spectrum of the $\rho$, $a_1$, $a_2$, and $f_0$ mesons and all their excitations, given in Table \ref{tab:inputs}. The mass eigenvalues are obtained by solving the Schr\"odinger-like equation for the bulk modes using a finite-difference discretization, where the spatial coordinate is divided into N points and the second derivative is approximated by a central difference scheme. This leads to a matrix eigenvalue problem, which can often be cast into a real symmetric form and solved using numerical eigenvalue solvers. By minimizing the difference between the predicted and experimental meson masses using gradient descent, we can optimize the neural networks to find the background geometry. The Python code, along with the trained models, is available at~\cite{RepoName}. The code can be easily modified to train further models, while the trained networks can be reused to explore additional observables and extensions of the AdS/QCD setup.
In contrast with the AdS/DL framework, the neural network model we construct here closely follows the ``deconstruction"~\cite{Arkani-Hamed:2001kyx,Cheng:2001nh,Abe:2002rj,Randall:2002qr,Falkowski:2002cm,Son:2003et,deBlas:2006fz,Erlich:2006hq} of a warped 5-dimensional theory with Dirichlet boundary conditions~\cite{Nakai:2014iea}.

For training, we have used the data of the mean values of the estimated Breit-Wigner (BW) masses from the PDG Summary Table \cite{ParticleDataGroup:2024cfk}. The Summary table estimates the $f_0(500)$ BW mass to be in the range 0.4 to 0.8 GeV, and the T-matrix pole ($\sqrt{s}$) mass to be in the range 0.40 to 0.55 GeV. To train the model, we have used the central value (0.475 GeV) of the T-Matrix Pole for this state. We scale down the loss associated with this mass to reduce its contribution to the total loss. This allows for greater variation in its mass during training. 
The estimated BW mass of $f_0(1370)$ as given in the Summary Table is in the range 1.2 to 1.5 GeV. We do not include this state in the training data and leave it for prediction by our trained model. 

Our results show good agreement with the experimental values for the $\rho$, $a_1$, $a_2$, and $f_0$ meson masses used in training. The mass of $f_0(1370)$, which is not well determined experimentally, was predicted to be around $1.49 \pm 0.02$ GeV, which is consistent with the experimental range of $1.200 \text{ to } 1.500$ GeV. We further predict the masses of the higher excited states of the $\rho$ and $f_0$ mesons, that were not used in training (not included in the Summary Table), whose results are given in Table \ref{tab:mass_spectra}. To assess the generalization capability of the trained model to other meson sectors, we examine its predictions for the excited-state masses of $\pi$ mesons. While the $\pi$ sector is not included in the training loss function, it is used as an additional constraint during hyperparameter optimization. The resulting predictions are in reasonable agreement with experimental values, indicating that the model captures key features of the underlying AdS/QCD dynamics.
In contrast with the GKK model \cite{Gherghetta_2009}, the dilaton profile we compute remains positive throughout the 5-dimensional space, which is consistent with the null energy condition~\cite{KIRITSIS200767, Karch:2010eg,Chelabi_2016}. The dilaton profile we obtain increases more steeply than a quadratic function. This finding is crucial because, though string-theoretical models predict a linear dilaton profile, such a profile would yield a continuum spectrum for glueballs~\cite{U_Gursoy_2008}. The dilaton wave profile, along with the scalar vevs, provides a deep enough potential for the pion wavefunction to support a $\sim 135$ MeV light state. 

In Section \ref{sec:ads_qcd}, we briefly introduce the bottom-up AdS/QCD, including the scalar, pseudo-scalar, vector, axial-vector, and higher spin mesons. In Section \ref{sec:methodology}, we discuss the methodology and define the Loss function for the neural network. Subsequently, in Section \ref{sec:results}, we present the numerical results and predictions, and conclude in Section \ref{sec:conclusion}. 
\section{Review of AdS/QCD}
\label{sec:ads_qcd}
The AdS/QCD model is based on the idea that the strong coupling regime of QCD can be described by a classical gravitational theory in a higher-dimensional AdS space. The correspondence between the two theories allows us to study non-perturbative aspects of QCD using classical gravity calculations in AdS space. The fields in the AdS space correspond to operators in the gauge theory, and the dynamics of these fields can be used to compute correlation functions and other observables in the gauge theory.

The metric in the 5-D AdS space is given by,
\begin{equation}
    g_{MN} dx^M dx^N = e^{2A(z)} \left(dz^2 + \eta_{\mu\nu} dx^\mu dx^\nu\right),
\end{equation}
where $z$ is defined as a dimensionless parameter related to the physical scale up to the radius of the AdS space ($L$) and the Minkowski metric is chosen to be $\eta_{\mu \nu} = \text{diag}(-1,+1,+1,+1)$. 

The 5D action, in this geometry, with a bulk scalar field $X$ transforming under the $SU(2)_L \times SU(2)_R$ symmetry is given by,
\begin{equation}
    I = -\int d^5x \sqrt{-g} e^{-\phi(z)} \Tr{\left[ |DX|^2 + m_X^2 |X|^2 - V(X) + \frac{1}{4g_5^2}(F_L^2 + F_R^2) \right]},
\end{equation}
where, $m_X^2 = -3$, $g_5^2 = 12 \pi^2/N_c$ ($N_c$ being the number of colors), $D^M X = \partial^M X - iA_L^M X + i X A_R^M$ is the covariant derivative, and $F_{L, R}^{M N} = \partial^M A_{L, R}^N - \partial^N A_{L, R}^M - i\left[A_{L, R}^M, A_{L, R}^N\right]$ are the field tensors. The scalar potential is taken to be $V(X) = \frac{4}{3}k_1 X^3 + 2k_2 X^4$.
 

The scalar field $X$ upon getting the VEV,
\begin{equation}
    \langle X \rangle \equiv \frac{v(z)}{2} \left(
    \begin{matrix}
        1 & 0 \\
        0 & 1
    \end{matrix} 
    \right) \ ,
\end{equation}
breaks the chiral symmetry $SU(2)_L \times SU(2)_R \rightarrow SU(2)_V$.

Varying the action, the dependence of VEV on the 5-dimensional space is given by,
\begin{equation}
    \partial_z \left(e^{-\phi(z) + 3A(z)} \partial_z v(z)\right) - e^{-\phi(z) + 5A(z)}\left(m_X^2 v(z) - V_{k}\left(v\right)\right) = 0,
    \label{eq:de_vev}
\end{equation}
where, $V_{k}(v(z)) = \frac{\partial }{\partial v(z)} \text{Tr} \left(V\left( \langle X \rangle \right) \right)=k_1 v(z)^2 + k_2 v(z)^3$. 

Demanding asymptotic AdS in the UV limit, the warp factor becomes $A(z)|_{z \rightarrow 0} = -\log(z)$ and the behavior of the VEV is given by,
\begin{equation}
    v(z\rightarrow0) = \alpha z + \beta z^3 \ ,
    \label{eq:vev_uv}
\end{equation}
where,
\begin{equation}
    \alpha = \zeta m_q L, \quad \beta = \frac{\Sigma}{\zeta} L^3.
    \label{eq:v_z_UV_behavior}
\end{equation}
Here, $m_q$ is the quark mass, $\Sigma$ is the chiral condensate, and $\zeta = \sqrt{N_c}/(2\pi)$ is a normalization factor \cite{PhysRevC.79.045203}. The Gell-Mann-Oakes-Renner (GMOR) relation becomes,
\begin{equation}
    f_\pi^2 m_\pi^2 = 2 m_q \Sigma \ .
    \label{eq:GMOR}
\end{equation}

\subsection{Meson Mass Spectrum}
\subsubsection{Scalar and Pseudo-Scalar Mesons}
The fluctuations of the bulk scalar field $X$ about the VEV could be written as $X(x, z) \equiv \left( v(z)/2 + S(x, z) \right) e^{2i\pi(x, z)}$, where $ S(x, z)$ and $\pi(x, z)$ are the scalar and the pion fields respectively. And expanding the fluctuation $S$ in its Kaluza-Klein modes (KK) $S(x, z) = \sum_n\mathcal{S}_n(x) S_n(z)$, the equation of motion of $S_n$ becomes,
\begin{equation}
    \partial_z \left(e^{-B_2(z)} \partial_z S_n(z)\right) - e^{-B_2(z)} e^{2A(z)} \left( m_X^2 - (2 k_1 v(z) + 3 k_2 v^2(z)) \right) S_n(z) = -e^{-B_2(z)} m^2_{f_0, n} S_n(z),
\end{equation}
where, $B_2(z) = \phi(z) - 3A(z)$, and $m_{f_0, n}^2$ are the discrete eigenvalues of the Sturm-Liouville problem. These discrete values of $m_{f_0, n}$ correspond to the mass of the scalar mesons.
This equation can be transformed into a Schr\"odinger equation, using the substitution $S_n(z) = e^{B_2(z)/2} \psi_{f_0,n}(z)$, giving,
\begin{equation}
    -\partial_z^2 \psi_{f_0,n} + V_{f_0}(z) \psi_{f_0,n} = m^2_{f_0,n} \psi_{f_0,n} \ ,
    \label{eq:schrodinger_f0}
\end{equation}
with the effective potential $ V_{f_0}(z)$ given by, 
\begin{equation}
    V_{f_0} = \left[ \frac{1}{4} \left( \partial_z B_2 \right)^2 - \frac{1}{2} \partial_z^2 B_2 + e^{2A(z)} \left( m_X^2 - (2 k_1 v(z) + 3 k_2 v^2(z)) \right) \right].
    \label{eq:potential_f0}
\end{equation}
Upon investigating the UV behavior of the potential $V_{f_0}$, we observe that $V_{f_0}(z \rightarrow 0) \rightarrow +\infty$, which ensures that the wave-function $\psi_{f_0,n}(z)$ goes to zero at the UV boundary, along with the normalizability condition $\psi_{f_0,n}(z\rightarrow\infty) \rightarrow 0$.

Varying the action with respect to the pion field $\pi(x, z)$, and expanding it in its KK modes, $\pi(x, z) = \sum_n \Pi_n(x)\pi_n(z)$, the equation of motion becomes,
\begin{equation}
    \frac{e^{B_2(z)}}{v^2(z)} \partial_z \left( v^2(z) e^{-B_2(z)} \partial_z \pi_n\right) + m^2_{\pi,n} (\pi_n - \xi_n) = 0 \ .
    \label{eq:pioneom}
\end{equation}
Here, we work in the gauge $A_5 = 0$, where the longitudinal component of the axial vector field, $A_\mu = (A_{L\mu} - A_{R \mu})/2 = A_{\perp \mu} + \partial_\mu \xi(x, z)$, is denoted by $\xi(x,z)=\sum_n \Xi_n(x)\xi_n(z)$, where $n$ denotes the KK mode. 

Also, varying the action with respect to the longitudinal component of the axial vector field, $\xi_n(z)$, gives the equation of motion,
\begin{equation}
    \partial_z \left( e^{-B_1(z)} \partial_z \xi_n \right) + g_5^2 v^2(z) e^{-B_2(z)} (\pi_n - \xi_n) = 0,
    \label{eq:xieom}
\end{equation}
where $B_1(z) = \phi(z) - A(z)$. To solve these coupled differential equations, it is possible to combine Eq. \ref{eq:pioneom} and Eq. \ref{eq:xieom}, into a Schr\"odinger-like equation \cite{PhysRevD.81.014024}, but doing so eliminates information from the mass spectrum \cite{PhysRevD.83.016002,PhysRevD.102.026013}. Specifically, we will only be able to obtain the mass eigenvalues of the excited states, losing information about the ground-state mass. We therefore choose to solve these two coupled second-order differential equations directly. Simplifying these equations, we get,
\begin{equation}
    \begin{split}
        \partial_z^2 \pi_n + \omega(z) \partial_z \pi_n + m_{\pi,n}^2 (\pi_n - \xi_n) & = 0, \\
        \partial_z^2 \xi_n - (\partial_z B_1(z)) \partial_z \xi_n + C(z) (\pi_n - \xi_n) & = 0,
    \end{split}
    \label{eq:pi_mesons_equation}
\end{equation}
where $\omega(z) = 2 \partial_z v(z)/v(z) - \partial_z B_2(z)$ and $C(z) = g_5^2 v^2(z) e^{2A(z)}$. These two equations can be solved with the boundary conditions $\pi_n = \xi_n = 0$ at $z \rightarrow 0$, and $\partial_z \pi_n = \partial_z \xi_n = 0$ at $z \rightarrow \infty$ to obtain the mass spectrum of the pseudo-scalar mesons.

\subsubsection{Vector and Axial-Vector Mesons}
For the vector mesons, $V_\mu = (A_{L\mu} + A_{R\mu})/2$, fixing the gauge $V_5 = 0$, the transverse component of the vector field can be expanded in terms of its KK modes as $V_{\mu \perp}^n (x, z) = \sum_n \mathcal{V}_\mu^{(n)}(x) V_n(z)$. Now, the equation of motion of $V_n(z)$ becomes,
\begin{equation}
    \partial_z \left( e^{-B_1(z)} \partial_z V_n(z) \right) + m_{\rho,n}^2 e^{-B_1(z)} V_n(z) = 0,
\end{equation}
where $B_1(z) = \phi(z) - A(z)$, and $m_{\rho,n}^2$ are the mass eigenvalues corresponding to the masses of the vector meson. To bring this equation to a Schr\"odinger-like form we substitute, $V_n(z) = e^{B_1(z)/2} \psi_{\rho,n}(z)$ and obtain,
\begin{equation}
    -\partial_z^2 \psi_{\rho,n}(z) + V_{\rho} \psi_{\rho,n}(z) = m_{\rho,n}^2 \psi_{\rho,n}(z),
    \label{eq:schrodinger_rho}
\end{equation}
where the effective potential is given by,
\begin{equation}
    V_{\rho} = \left[ \frac{1}{4} \left( \partial_z B_1 \right)^2 - \frac{1}{2} \partial_z^2 B_1 \right].
    \label{eq:potential_rho}
\end{equation}
The masses of the vector meson are computed by demanding the Dirichlet boundary condition $\psi_{\rho,n} \rightarrow 0$ at UV and for large $z$. 

Similarly, for the axial-vector mesons, choosing the gauge $A_5 = 0$, and expanding the axial vector field in its KK decomposition, $A_{\mu \perp}^n (x, z) = \mathcal{A}_\mu^{(n)}(x) a_n(z)$, we get the equation of motion for $a_n(z)$ as,
\begin{equation}
    \partial_z(e^{-B_1(z)} \partial_z a_n) + \left(m_{a_1, n}^2 - g_5^2 e^{2A(z)} v^2(z)\right) e^{-B_1(z)} a_n = 0.
\end{equation}
Here we denote the axial-vector meson masses by $m_{a_1, n}^2$. Substituting $a_n(z) = e^{B_1(z)/2} \psi_{a_1, n}(z)$, we obtain,
\begin{equation}
    -\partial_z^2 \psi_{a_1, n}(z) + V_{a_1} \psi_{a_1, n}(z) = m_{a_1, n}^2 \psi_{a_1, n}(z),
    \label{eq:schrodinger_a1}
\end{equation}
such that $\psi_{a_1, n}(z)$ satisfies Dirichlet boundary condition at UV and at large $z$. The effective potential is given by,
\begin{equation}
    V_{a_1} = \left[ \frac{1}{4} \left( \partial_z B_1 \right)^2 - \frac{1}{2} \partial_z^2 B_1 + g_5^2 e^{2A(z)} v^2(z) \right].
    \label{eq:potential_a1}
\end{equation}

\subsubsection{Higher-Spin Mesons}
To study the higher spin mesons, consider a rank-G symmetric tensor field, and following the same procedure as before, we can show that the z-dependent part of the 5-D bulk tensor field ($G(x,z)=\sum_n \mathcal{G}_n (x)G_n(z)$) follows the equation of motion,
\begin{equation}
    \partial_z \left( e^{-B_G(z)} \partial_z G_n(z) \right) + m_{a_G,n}^2 e^{-B_G(z)} G_n(z) = 0,
\end{equation}
where $B_G(z) = \phi(z) - (2G - 1) A(z)$. For a spin-2 meson like the $a_2$ meson, we have $G = 2$ and substituting $G_n(z) = e^{B_G(z)/2} \psi_{a_2, n}(z)$, we obtain the Schr\"odinger-like equation,
\begin{equation}
    -\partial_z^2 \psi_{a_2, n}(z) + V_{a_2} \psi_{a_2, n}(z) = m_{a_2, n}^2 \psi_{a_2, n}(z) \ ,
    \label{eq:schrodinger_a2}
\end{equation}
where the effective potential is given by, 
\begin{equation}
    V_{a_2} = \left[ \frac{1}{4} \left( \partial_z B_2 \right)^2 - \frac{1}{2} \partial_z^2 B_2 \right].
    \label{eq:potential_a2}
\end{equation}
with $B_2(z) = \phi(z) - 3A(z)$. The mass eigenvalues $m_{a_2, n}^2$ of the spin-2 meson are obtained by solving the equation of motion and imposing the Dirichlet boundary conditions on the wave functions.

With these, in the next section we discuss a neural network architecture to solve the Schr\"odinger-like equations, along with the boundary conditions, to obtain the warp factor, dilaton profile, and the chiral symmetry breaking potential by training the model with the masses $m_{f_0,n}$, $m_{\rho, n}$, $m_{a_1, n}$ and $m_{a_2, n}$ in Table \ref{tab:inputs}. Moreover, the network will be used to predict the pion mass spectrum, $m_{\pi,n}$, and compare them with experimental values.

\section{Methodology}
\label{sec:methodology}
\subsection{Model Architecture}
We model the warp factor $A(z)$ and the scalar VEV $v(z)$ using a trainable neural network. We take two feed-forward neural networks with input $z \in (0, z_{max})$, and outputs a scalar value $A_{\text{NN}}(z)$ and $v_{\text{NN}}(z)$ respectively. The value of $z_{max}$ is chosen sufficiently large such that boundary conditions on wavefunctions are satisfied.
We employ a transformation function on the output of the neural network to get the approximated functions for the warp factor $A(z)$ and $v(z)$, defined as
\begin{eqnarray}
        A(z) &= -\ln(z) + \frac{z}{10} A_{\text{NN}}(z) \ , \nonumber \\
        v(z) &= \frac{\alpha z + \beta z^3 + v_{\text{NN}}(z) z^4}{1 + z^3} \ ,
    \label{eq:hard_boundary}
\end{eqnarray}
This transformation ensures the correct boundary condition at the UV region.
The denominator in $v(z)$ is introduced in order to prevent numerical instability in the IR region. This does not affect the UV behavior of v(z) up to order of $\mathcal{O}(z^4)$.

While training, the chiral condensate ($\Sigma$) is made sure to lie within a reasonable range ($0.01 \ GeV^3 : 0.03 \ GeV^3$), and is defined in terms of the training parameter $\theta$ as,
\begin{equation}
    \Sigma = a + (b - a)\sigma(\theta)
    \label{eq:sigma_theta_rel}
\end{equation}
where, $\sigma$ is the sigmoid function and $a = 0.01$ GeV$^3$ and $b = 0.03$ GeV$^3$. The quark mass $m_q$ is then computed using the GMOR relation given in Eq. \ref{eq:GMOR}, and this is, in turn, used to compute the $\alpha$ and $\beta$ values from Eq. \ref{eq:v_z_UV_behavior}.

For a given background $A(z)$ and the VEV $v(z)$, the $B_2(z)$ could be computed from Eq. \ref{eq:de_vev} as,
\begin{equation}
    \partial_z B_2(z) = \frac{1}{\partial_z v(z)} \left( \partial^2_z v(z) + e^{2A(z)} (3v(z) + V_k(v(z))) \right),
    \label{eq:B2_dash}
\end{equation}
where, $m_X^2 = -3$ and,
\begin{equation}
    V_k(v(z)) = k_1 v(z)^2 + k_2 v(z)^3 \ .
\end{equation}
The coefficients $k_1$ and $k_2$ are the trainable parameters. 
We have observed that adding a cubic interaction term along with a quartic term significantly reduces the loss. This may indicate either insufficient hyperparameter tuning in the quartic-only case or that a cubic interaction is genuinely required in the holographic model.

With $\partial_z B_2(z)$ computed in Eq.~\ref{eq:B2_dash}, we obtain,
\begin{equation}
\partial_z B_1(z) = \partial_z B_2(z) + 2\,\partial_z A(z), \qquad
\partial_z \phi(z) = \partial_z B_2(z) + 3\,\partial_z A(z).
\end{equation}
Using these relations, the effective potentials appearing in Eq.~\ref{eq:potential_f0}, Eq. \ref{eq:potential_rho}, Eq. \ref{eq:potential_a1}, and Eq. \ref{eq:potential_a2} are evaluated. The resulting Schr\"odinger-like equations are then solved numerically using the finite-difference method, with boundary conditions $\psi_n(0)=0$ and $\psi_n(z_{\mathrm{max}})=0$, yielding the meson mass eigenvalues. 

The AdS radius $L$ is determined in terms of the trainable parameter $l$ as,
\begin{equation}
    L = \log(e + l^2) \ ,
    \label{eq:ads_radius}
\end{equation}
to ensure that its value remains greater than 1, while also preventing it from taking excessively large values, during training. 

The values of $k_1$ and $k_2$ significantly influence the potentials and, consequently, the mass spectrum. Choosing a good initial guess for these parameters is crucial for the convergence of the training. The initial values of the trainable parameters $k_1$ and $k_2$ are chosen using a Bayesian optimization algorithm called the Tree-Structured Parzen Estimator (TPE) \cite{10.1145/3377930.3389817} from the Optuna library \cite{optuna_2019} (See Appendix \ref{appendix:training_details} for more details). The initial values of the other two trainable parameters are taken as $\theta = 0.0$ and $l = 1.0$. 
With the two neural networks for $A(z)$ and $v(z)$, and the four trainable parameters $\theta$, $k_1$, $k_2$, and $l$, we aim to fit the mass spectra of scalar ($f_0$), vector ($\rho$), axial-vector ($a_1$) and spin-2 ($a_2$) mesons, and use the trained model to predict the masses of the pseudo-scalar ($\pi$) mesons by solving Eq. \ref{eq:pi_mesons_equation}. 
The mass eigenvalues are calculated using the finite-difference method by writing the discretized form of the Schrödinger-like equations in matrix form. This discretized form is constructed by discretizing the z-coordinate (See Appendix \ref{appendix:numerical_routine}). The mass eigenvalues of the Schrödinger-like equations are directly controlled by the value of the effective potential at these discretized points, which appear as the diagonal elements of the matrix. Therefore, the model trains the values of the functions $A(z)$ and $v(z)$ at these discretized points, thereby learning the effective potential which can yield the correct mass eigenvalue.

\subsection{Loss Functions}
\label{sec:loss_functions}

To get the mass eigenvalues, we solve the Schr\"odinger-like equations using the finite-difference method. We discretize the spatial coordinate $z$ into N points ranging from $z = 0$ to $z = z_{max}$ with a step size of $h$. The discretized form of the Schr\"odinger-like equation $-\partial_z^2 \psi(z) + V_{\mu}(z) \psi(z) = \lambda_{\mu} \psi(z)$ can be written in the form of a matrix eigenvalue problem (See Appendix \ref{appendix:numerical_routine}), which can be solved using numerical eigenvalue solvers to obtain the non-dimensional mass eigenvalues $\lambda_{\mu}$ for the meson type $\mu \in \{\rho, a_1, a_2, f_0\}$. The Dirichlet boundary conditions $\psi(0) = 0$ and $\psi(z_{max}) = 0$ result in a real symmetric tri-diagonal matrix whose eigenvalues can be efficiently computed using the \texttt{torch.linalg.eigvalsh} function in PyTorch, which under the hood uses the LAPACK library (or its GPU analogs). It returns the eigenvalues in ascending order. To convert them to physical units, the eigenvalues are divided by $L^2$,
\begin{equation}
    m_{\mu, n}^2 = \frac{\lambda_{\mu, n}}{L^2}
\end{equation}
where, $L$ is the AdS radius and $\mu$ index identifies the meson type (scalar $f_0$, axial vector $a_1$, vector $\rho$ or tensor $a_2$). The error for the $n^{th}$ mass of the meson is given by,
\begin{equation}
    E_{\mu, n} = \frac{|m_{\mu, n}^2 - \omega_{\mu,n}^2|}{\omega_{\mu,n}^2},
    \label{eq:mass_error}
\end{equation}
where $\omega_n$ are the experimental values of the meson masses as given in \cite{ParticleDataGroup:2024cfk}. 
The Loss function is then defined as,
\begin{equation}
    \mathcal{L}_{\text{mass}}^{(\mu)} = \frac{1}{N_{\mu}} \sum_{n=1}^{N_{\mu}}  \Lambda_{\mu, n} E_{\mu, n},
\end{equation}
where $N_{\mu}$ is the number of excited mesons considered for each kind, and $\Lambda_{\mu, n}$ is a hyperparameter that controls the weightage of each meson mass in the total loss function. We have employed an adaptive weighting scheme where $\Lambda_{\mu, n}$ at each training step is proportional to the loss contributed by that particular mass (See Appendix \ref{appendix:training_details}: Eq. \ref{eq:Lambda_mass}). This ensures that the model focuses more on the masses that are harder to fit. Further, we have scaled the error for the first mass of the $f_0 (500)$ meson $E_{f_0,1}$ by a factor of 0.1, as it is an outlier in the mass spectrum and the experimental value is not well determined.
The total mass loss function used in the training is given by
\begin{equation}
    \mathcal{L}_{\text{mass}} = \mathcal{L}_{\text{mass}}^{(\rho)} + \mathcal{L}_{\text{mass}}^{(a_1)} + \mathcal{L}_{\text{mass}}^{(a_2)} + \mathcal{L}_{\text{mass}}^{(f_0)}.
\end{equation}
It is essential for the Loss function to be differentiable so that the gradients can be calculated using backpropagation. Although the internal algorithm (LAPACK) itself is not automatically differentiable, the PyTorch library has implemented a mathematical formula that is pre-programmed into it. (See Appendix \ref{appendix:numerical_routine} for more details.)

The equation for $\partial_z B_2$ given in Eq. \ref{eq:B2_dash} can become numerically unstable if $\partial_z v(z) = 0$. Since $\partial_z v(z)|_{z=0} = \alpha$, and $\alpha$ is positive, we demand that $\partial_z v(z) > 0$ for all $z$. This ensures that $\partial_z v(z)$ never crosses zero, therefore avoiding numerical instabilities. To enforce this condition, we add another penalty term to the loss function given by
\begin{equation}
    \mathcal{L}_{v'} = \text{ReLU}\left(-\partial_z v(z)\right),
    \label{eq:pos_v_dash_loss}
\end{equation}
where ReLU is the rectified linear unit function defined as $\text{ReLU}(x) = \max(0, x)$. This term penalizes any negative values of $\partial_z v(z)$.

\begin{figure}
    \centering
    \includegraphics[width=0.95\textwidth]{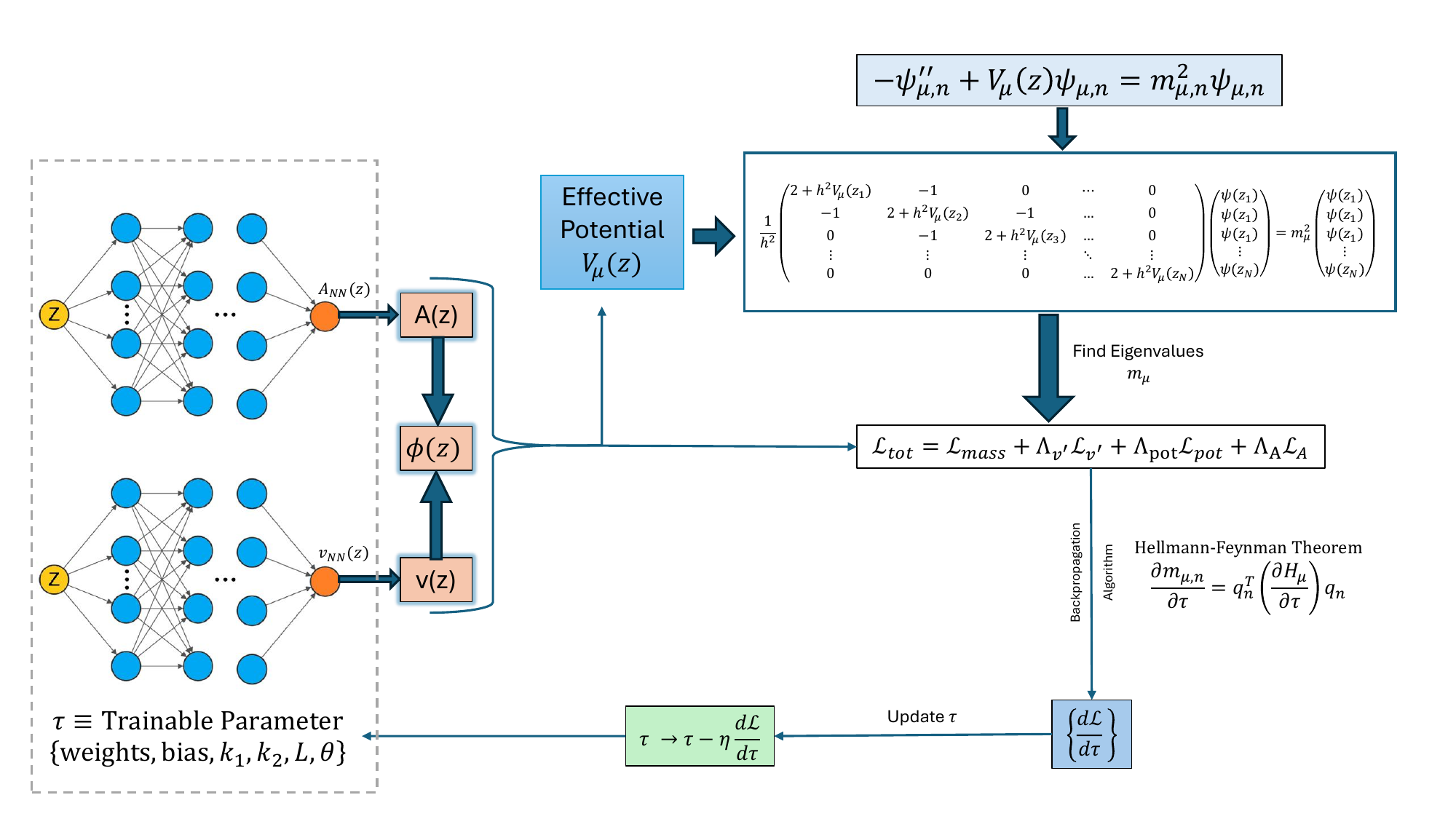}
    \caption{Summary of the Neural Network training architecture.}
    \label{fig:model_architecture}
\end{figure}

To confine the wavefunction within the AdS space and ensure normalizability of the wavefunction at large $z$, we expect the potentials $V_{\mu}(z)$ to be non-decreasing functions of $z$ in the IR region. This can be imposed by demanding that $\partial_z V_{\rho}(z)$ and $\partial_z V_{a_2}(z)$ are positive at large $z$. This automatically also ensures that the potentials given in Eq.  \ref{eq:potential_f0} and Eq. \ref{eq:potential_a1} are also confining. To impose this condition, we add another penalty term to the loss function given by
\begin{equation}
    \mathcal{L}_{\text{pot}} = \text{ReLU}\left(-\partial_z V_{\rho}(z)\right) + \text{ReLU}\left(-\partial_z V_{a_2}(z)\right),
    \label{eq:pos_pot_dash_loss}
\end{equation}
where $z$ is chosen randomly at each training step at a sufficiently large value close to $z_{max}$. We expect $A(z)$ to be negative and decreasing at the IR scale due to warped geometry. 
This is imposed by adding another penalty term,
\begin{equation}
    \mathcal{L}_{A} = \text{ReLU}\left(A(z_{max}) + \ln(z_{max}) - \Delta\right) + \text{ReLU}\left(\partial_z A(z_{max})\right),
 \end{equation}
where $\Delta$ is a constant we choose to be 1.5. The total loss function now becomes,
\begin{equation}
    \mathcal{L}_{\text{tot}} = \mathcal{L}_{\text{mass}} + \Lambda_{v'} \mathcal{L}_{v'} + \Lambda_{\text{pot}} \mathcal{L}_{\text{pot}} + \Lambda_A \mathcal{L}_A,
    \label{eq:loss_total}
\end{equation}
where $\Lambda_{v'}$, $\Lambda_{\text{pot}}$, and $\Lambda_A$ are hyperparameters that control the weightage of the penalty terms. The model is trained to minimize the total loss function $\mathcal{L}_{\text{tot}}$ by optimizing the neural network's weights and biases, as well as the trainable parameters $\theta$, $k_1$, $k_2$, and $l$. A schematic diagram summarizing the entire architecture is shown in Figure \ref{fig:model_architecture}.

\section{Numerical Results}
\label{sec:results}
We perform a hyperparameter search for the initial values of $k_1$ and $k_2$ using Optuna's built-in TPE sampler, in the range $[-20, 20]$ (See Appendix \ref{appendix:training_details}). We observe that the model converges to a lower loss value for negative $k_1$ and positive $k_2$, resulting in a bounded scalar field potential. The best initial guess was found to be around $k_1 = -2.2$ and $k_2 = 12.0$. With this initial guess, we train our model several times and select only those runs in which the total loss converges to a value below $0.01$. We have selected 10 such runs, whose training loss and the trained model parameters ($k_1$, $k_2$, $L$, $\theta$) corresponding to each run are listed in Table \ref{tab:training_loss}. Further details of the training procedure, along with the hyperparameters used, are provided in Appendix \ref{appendix:training_details}.
\begin{figure}
    \centering
    \includegraphics[width=0.8\textwidth]{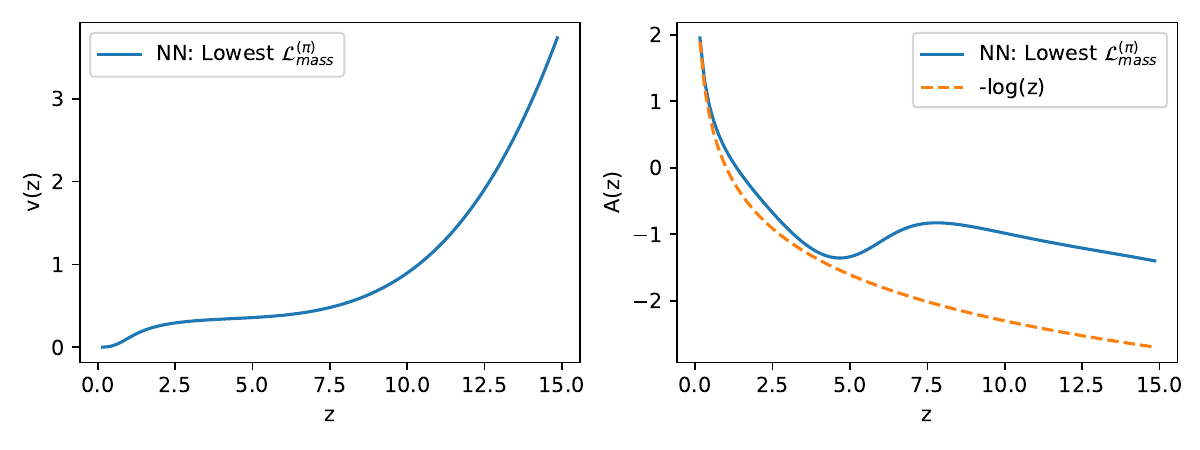}
    \caption{The learned functions $v(z)$ (left) and $A(z)$ (right) after training corresponding to the best fit to the $\pi$ meson masses.}
    \label{fig:v_A_plots}
\end{figure}

\begin{table}[!ht]
\centering
\begin{tabular}{|l|c|l|c|}
\hline
$L$ & $1.234 \pm 0.229 \  GeV^{-1}$ & $\theta$ & $3.697 \pm 1.305$ \\
$k_1$ & $-4.332 \pm 1.022$ & $k_2$ & $8.600 \pm 2.355$ \\
\hline
\end{tabular}
\caption{The trained model parameters $L = \log{(e + l^2)}$, $\theta$, $k_1$ and $k_2$. The values are given as mean $\pm$ standard deviation over different runs.}
\label{tab:model_parameters}
\end{table}

\begin{figure}
    \centering
    \includegraphics[width=0.8\textwidth]{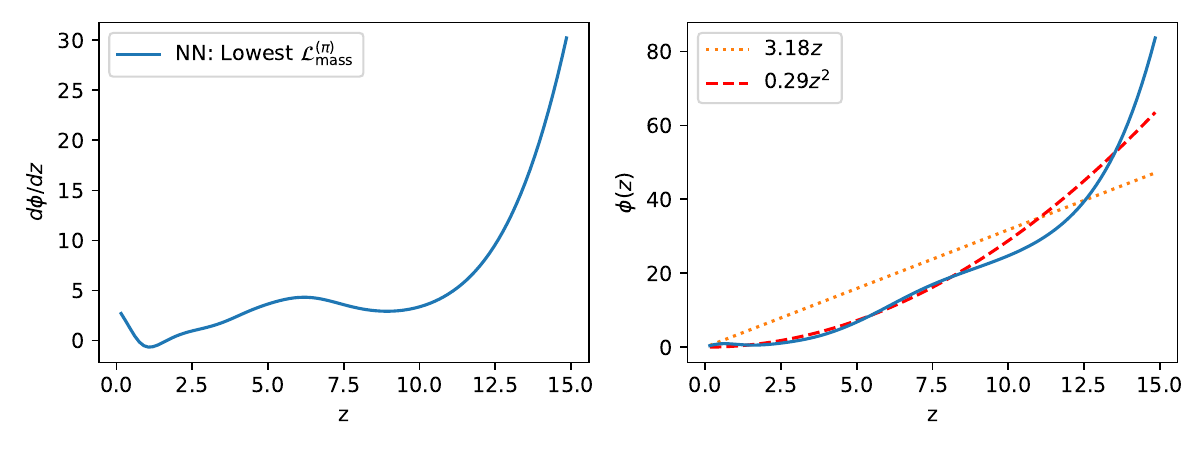}
    \caption{The plot of $d\phi/dz$ (left) and $\phi(z)$ (right) for the trained models corresponding to the best fit to the $\pi$ meson masses. The dotted line and the dashed line in the right panel represent a linear and quadratic fit to the mean value of $\phi(z)$, respectively.}
    \label{fig:phi_plot}
\end{figure}

\begin{figure}
    \centering
    \includegraphics[width=0.8\textwidth]{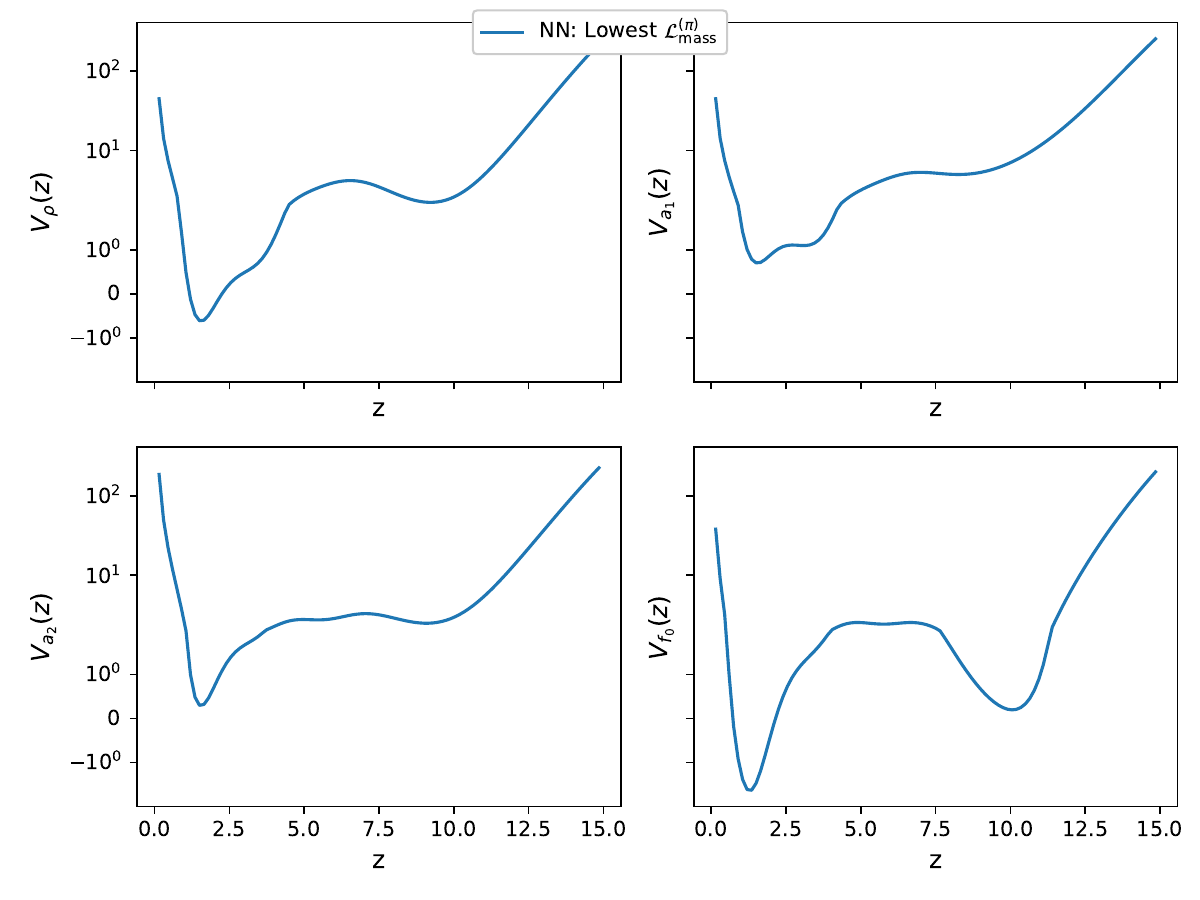}
    \caption{Effective potentials (log scaled) for the $\rho$, $a_1$, $a_2$, and $f_0$ mesons corresponding to the best fit to the $\pi$ meson masses. }

    \label{fig:potentials_plot}
\end{figure}

The trained functions $v(z)$ and $A(z)$ are shown in Fig. \ref{fig:v_A_plots}, and the mean of the model parameters obtained over different runs is given in Table \ref{tab:model_parameters}. The trained parameter $\theta$ can be used to get the value of $\Sigma$ from Eq. \ref{eq:sigma_theta_rel}, and use the GMOR relation in Eq. \ref{eq:GMOR} to get $m_q$ as,
\begin{equation}
    \Sigma = (0.3071 \pm 0.0042 \ \text{GeV})^3 \quad \text{and} \quad m_q = 0.0029 \pm 0.0001 \ \text{GeV},
\end{equation}
where the values are given as mean $\pm$ standard deviation over different runs. With the given $v(z)$ and $A(z)$, we can obtain $d\phi/dz$ which is plotted in Fig. \ref{fig:phi_plot} (left). Integrating $d\phi/dz$ gives us $\phi(z)$ up to an additive constant, also given in Fig. \ref{fig:phi_plot} (right). We have plotted a linear and quadratic fit to the function $\phi(z)$. Additionally, we plot the effective potentials for the $\rho$, $a_1$, $a_2$, and $f_0$ mesons in Fig. \ref{fig:potentials_plot}.

\begin{figure}[!ht]
    \centering

    \includegraphics[width=0.8\textwidth]{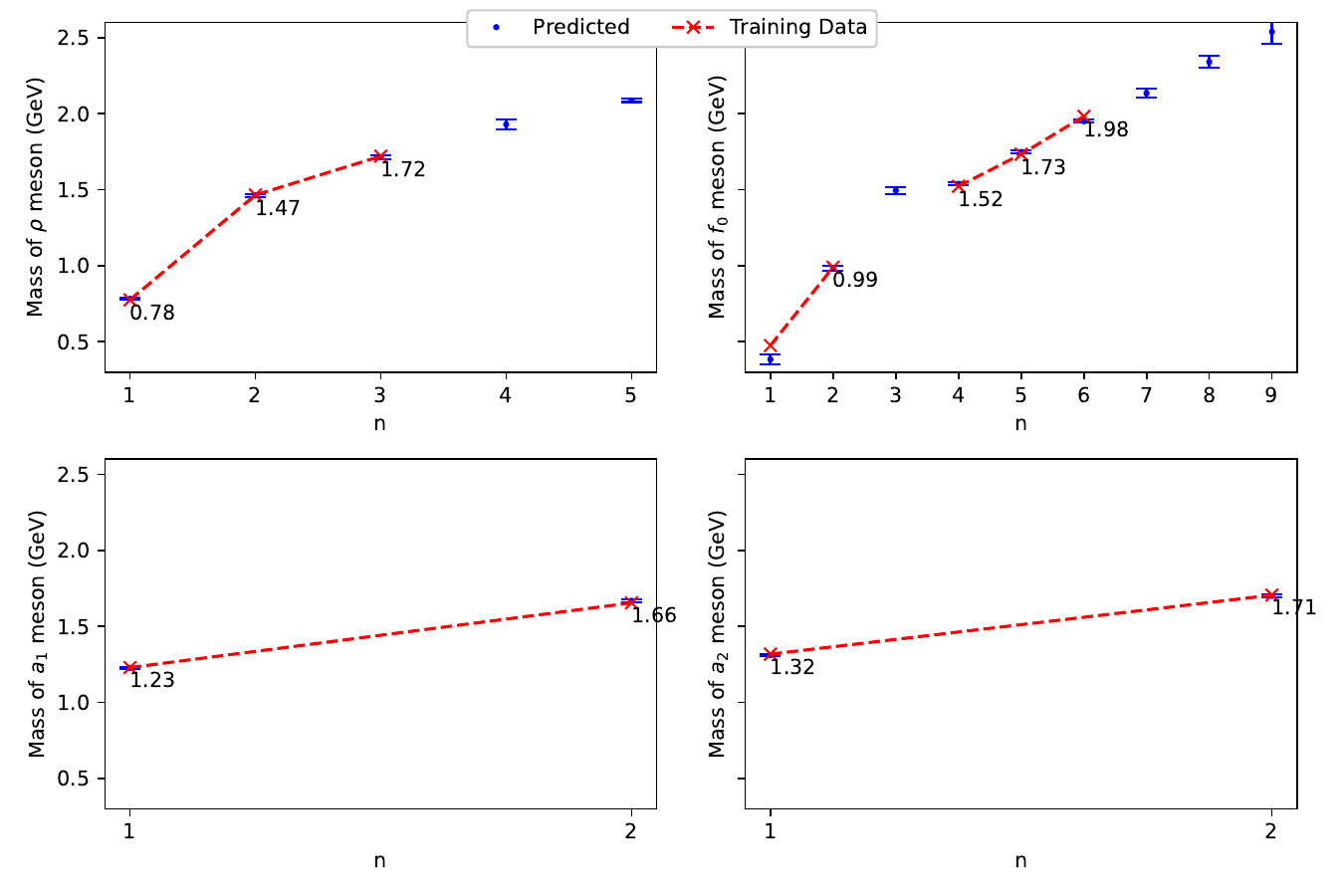}

    \caption{Mass spectra of the $\rho$, $f_0$, $a_1$ and $a_2$ mesons \cite{ParticleDataGroup:2024cfk}. The predicted masses correspond to the mean, and the error bars show the standard deviation across multiple runs.}

    \label{fig:mass_spectra}
\end{figure}

The predicted spectra of the trained masses of the $\rho$, $a_1$, $a_2$, and $f_0$ mesons obtained from our model are given in Table \ref{tab:mass_spectra} and are plotted with respect to n (radial excitation number) in Fig. \ref{fig:mass_spectra}. The predicted masses agree well with the experimental values. The comparatively larger error for the $f_0(500)$ is not unexpected, as it reflects the larger experimental uncertainty of this broad resonance (400 to 500 MeV), and the reduced weightage assigned to its loss during training, which allows for greater variation in its predicted mass. 
Using the trained model, we predict the mass of $f_0(1370)$, which we skipped in training due to the large uncertainty in its experimental value. The predicted mass of $f_0(1370)$ is $1.49 \pm 0.02$ GeV, which is consistent with the given range of $1.2$ to $1.5$ GeV in the Summary Table of the Particle Data Group~\cite{ParticleDataGroup:2024cfk}. We further predict the higher excited states of the $\rho$ and $f_0$ mesons, which are not well established experimentally. The predicted masses of the $\rho(1900)$, $\rho(2150)$, $f_0(2200)$, $f_0(2330)$ and $f_0(2470)$ are given in Table \ref{tab:mass_spectra} and are within the expected range of experimental uncertainties. We have omitted the state $\rho(1570)$ in the table, because it may be an OZI-violating decay mode of $\rho(1700)$.

\begin{table}[!ht]
    \centering
    \begin{tabular}{|c|c|c|c|}
    \hline
    \textbf{Meson} & \textbf{State} & \textbf{Predicted Mass (GeV)} & \textbf{\% Error} \\
    \hline
    $\rho$  
        & $1$ & $0.78 \pm 0.01$ &  0.95\% \\
        & $2$ & $1.46 \pm 0.01$ &  0.15\% \\
        & $3$ & $1.71 \pm 0.01$ &  0.37\% \\
        & $4^\prime$ & $1.93 \pm 0.03$ &  1.04\% \\
        & $5^\prime$ & $2.09 \pm 0.01$ &  0.41\% \\
    \hline
    $a_1$  
        & $1$ & $1.23 \pm 0.01$ &  0.20\% \\
        & $2$ & $1.67 \pm 0.01$ &  0.88\% \\
    \hline
    $a_2$  
        & $1$ & $1.31 \pm 0.00$ &  0.49\% \\
        & $2$ & $1.70 \pm 0.01$ &  0.33\% \\
    \hline
    $f_0$  
        & $1$ & $0.38 \pm 0.03$ &  19.18\% \\
        & $2$ & $0.98 \pm 0.01$ &   0.74\% \\
        & $3^\prime$ & $1.49 \pm 0.02$ &   1.54\% \\
        & $4$ & $1.54 \pm 0.01$ &   1.07\% \\
        & $5$ & $1.75 \pm 0.01$ &  0.87\% \\
        & $6$ & $1.95 \pm 0.01$ &  1.62\% \\
        & $7^\prime$ & $2.13 \pm 0.03$ &  2.40\% \\
        & $8^\prime$ & $2.34 \pm 0.04$ &  0.01\% \\
        & $9^\prime$ & $2.54 \pm 0.08$ &  2.77\% \\
    \hline
    \end{tabular}
    \caption{Predicted meson masses (in GeV). The predictions are reported as the mean $\pm$ standard deviation across different runs. The \% Error is calculated using the mean value of the predicted mass. The unprimed states were used for model training.}
    \label{tab:mass_spectra}
\end{table}

\begin{table}[!ht]
\centering
\begin{tabular}{|c|c|c|c|}
\hline
\multirow{2}{*}{\textbf{State}} &
\multirow{2}{*}{\parbox{5cm}{\centering \textbf{Predicted Mass (GeV) \\ (Mean $\pm$ Std)}}} &
\multirow{2}{*}{\parbox{5cm}{\centering \textbf{Predicted Mass (GeV) \\ (Best fit: Lowest $\mathcal{L}_{\text{mass}}^{(\pi)}$)}}} &
\multirow{2}{*}{\parbox{2cm}{\centering \textbf{\% Error}}}
\\
& & & \\
\hline
1 & $0.157 \pm 0.017$ & 0.137 & 1.71 \\
2 & $1.720 \pm 0.017$ & 1.717 & 32.11 \\
3 & $1.921 \pm 0.023$ & 1.918 & 6.58 \\
\hline
\end{tabular}
\caption{Predicted $\pi$ meson masses (in GeV) \cite{ParticleDataGroup:2024cfk}. The predicted masses are reported as the mean $\pm$ standard deviation across different runs. The best fit values correspond to the run with lowest $\mathcal{L}_{\text{mass}}^{(\pi)}$ in Table \ref{tab:training_loss}. The percentage error is computed using the best-fit value.}
\label{tab:pi_masses}
\end{table}

To validate our model, we predict the masses of the pseudo-scalar meson $\pi$. We use the finite difference method to solve the coupled second-order differential equation given in Eq. \ref{eq:pi_mesons_equation} using the learned parameters and functions.
This time, we obtain a matrix for a generalized eigenvalue equation (see Appendix \ref{appendix:numerical_routine} for more details), which we solve using SciPy's built-in function \texttt{scipy.linalg.eigvals}.

\begin{figure}[!ht]
    \centering
    \includegraphics[width=0.5\textwidth]{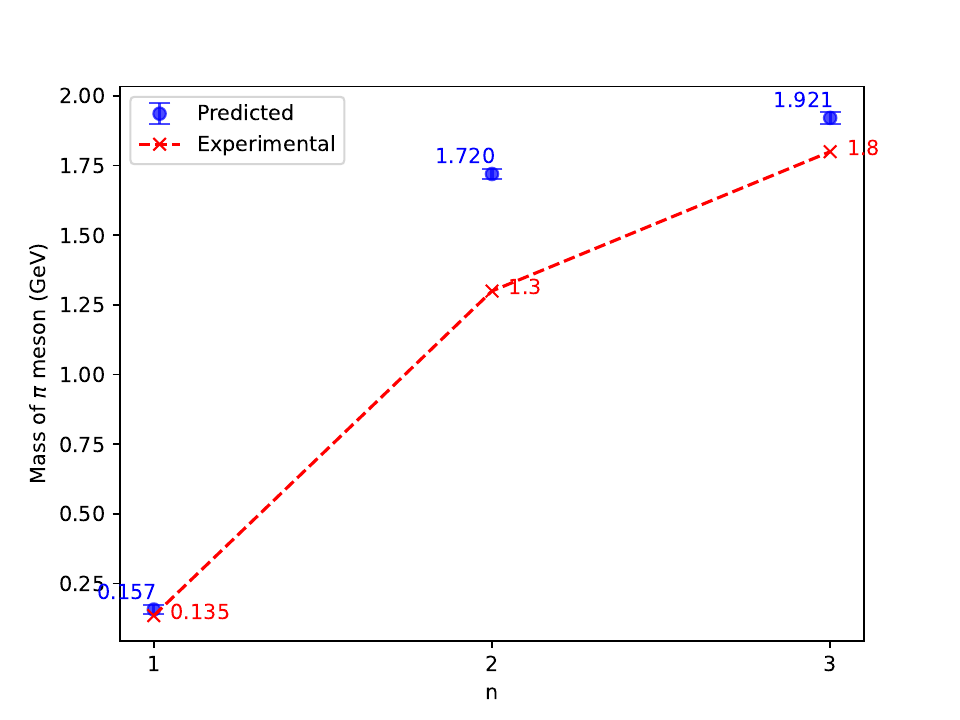}
    \caption{Mass spectrum of the predicted $\pi$ meson as shown in Table \ref{tab:pi_masses}. The error bars represent the standard deviation over different runs. 
    }
    \label{fig:pi_masses}
\end{figure}

The predicted mass eigenvalues $m_n$ obtained (after converting to physical units) using our trained model are given in Table \ref{tab:pi_masses} and plotted in Fig. \ref{fig:pi_masses}. Here, the best fit values are obtained for the run with the lowest value of $\mathcal{L}_{mass}^{(\pi)}$. The Loss corresponding to the $\pi$ meson masses for each run is given in Table \ref{tab:training_loss}. Although the $\pi$ meson masses are not included in the training loss, they are taken into account during the hyperparameter optimization procedure. The predicted values nevertheless show good agreement with experimental data, except for the first excited state, where the model significantly overshoots the mass. 

\section{Conclusion}
\label{sec:conclusion}

The bottom-up approach in holographic QCD explores the functional space of all possible background metrics and scalar potentials, searching for functions that best fit the QCD observables. With enough phenomenological and theoretical constraints, the bottom-up approach can severely restrict the allowed background functions and may select a class of viable models. We have provided a deep-learning architecture that searches in the possible functional space of the warp-factor $A(z)$, the dilaton profile $\phi(x)$, along with the possible values of cubic ($k_1$) and quartic ($k_2$) terms in the scalar potential, and the quark condensate $\Sigma$. This space is constrained by fitting the mass spectrum of the unflavored mesons ($\rho, a_1, a_2$ and $f_0$) and their excitations. 

The duality between AdS/QCD and deep neural networks, as shown in \cite{PhysRevD.98.046019}, raises intriguing questions about their physical origins and suggests a new methodology for investigating both frameworks. This was pioneered in~\cite{Akutagawa_2020}, where the authors trained for the masses of the $\rho$ and $a_1$ meson, through which the $A(z)$ and $\phi(z)$ were determined. They solved the differential equation of the KK mode using a discretized Euler-type scheme, which may accumulate numerical errors over long evolutions. The eigenvalues are calculated by traversing all the values of masses to search for the mass for which KK modes are normalizable. Since the axial sector was not studied, the information about the chiral symmetry breaking by the bulk scalar field is not well understood. Including higher-meson masses and the axial sector, along with the $f_0$ mesons, becomes computationally challenging because the network's structure is determined by the theory itself.

In contrast, we propose here to train the architecture using the finite-difference method to directly obtain the mass eigenvalue, thereby making it more efficient, less error-prone, and more extensible to higher excitation states of the meson tower. Additionally, the introduction of the scalar vev has enabled us to determine the masses of all unflavored mesons. This strategy is flexible and highly tunable, leaving scope to impose additional constraints on the functional space via additional loss functions. 

The strategy we use here is described in Appendix~\ref{appendix:numerical_routine}, where the eigenvalues and functions of the Schr\"odinger-like equations are computed by diagonalizing the matrix obtained from latticized 5-dimensional space. Unlike in~\cite{Akutagawa_2020}, here there is no emergent metric arising from the AdS/DL correspondence; instead, the framework closely follows the latticized 5D action via dimensional deconstruction~\cite{Arkani-Hamed:2001kyx,Cheng:2001nh,Abe:2002rj,Randall:2002qr,Falkowski:2002cm,Son:2003et,deBlas:2006fz,Erlich:2006hq,Nakai:2014iea} to obtain a linear moose model. The neural network enters the linear moose model at the prediction of the potential for the wavefunction. 

Though the $z^2$ behavior for the dilaton profile is anticipated by several phenomenological models~\cite{Karch:2006pv,Gherghetta_2009,Zhang_2010,PhysRevD.100.106009} for linear confinement, string theoretic models often assume a linear profile. In this work, the predicted dilaton, in Fig. \ref{fig:phi_plot}, exhibits a profile that increases more steeply than quadratic behavior, and is positive throughout the region, making it consistent with the null energy condition \cite{KIRITSIS200767}. Moreover, it predicted the pion mass and its excitations to a reasonable level of accuracy. As mentioned earlier, our numerical experiment suggests that for a more precise fitting of the mass spectra, a cubic interaction, in addition to the quartic interaction, is necessary.

\section*{Acknowledgments}
M.T.A. acknowledges the financial support of DST through the INSPIRE Faculty grant  DST/INSPIRE/04/2019/002507.
\appendix
\section*{Appendix}
\section{Numerical Routine}
\label{appendix:numerical_routine}

The discretized form of the Schr\"odinger-like equations $-\partial_z^2 \psi(z) + V_{\mu}(z) \psi(z) = \lambda_{\mu} \psi(z)$ for $\mu \in \{\rho, a_1, a_2, f_0\}$ is given by
\begin{equation}
    -\frac{\psi(z + h) - 2 \psi(z) + \psi(z - h)}{(h)^2} + V_{\mu}(z) \psi(z) = \lambda_{\mu} \psi(z),
\end{equation}
where $\lambda_{\mu}$ is the eigenvalue of the equation and $h$ is the step size. This can be written in the matrix form as
\begin{equation*}
    \frac{1}{h^2}
    \begin{pmatrix}
        2 + h^2 V_{\mu}(z_1) & -1 & 0 & \cdots & 0 \\
        -1 & 2 + h^2 V_{\mu}(z_2) & -1 & \cdots & 0 \\
        0 & -1 & 2 + h^2 V_{\mu}(z_3) & \cdots & 0 \\
        \vdots & \vdots & \vdots & \ddots & \vdots \\
        0 & 0 & 0 & \cdots & 2 + h^2 V_{\mu}(z_{N-1})
    \end{pmatrix}
    \begin{pmatrix}
        \psi(z_1) \\
        \psi(z_2) \\
        \psi(z_3) \\
        \vdots \\
        \psi(z_{N-1})
    \end{pmatrix}
    = \lambda_{\mu}
    \begin{pmatrix}
        \psi(z_1) \\
        \psi(z_2) \\
        \psi(z_3) \\
        \vdots \\
        \psi(z_{N-1})
    \end{pmatrix}
\end{equation*}
where $z_i = z_0 + i h$ for $i = 0, 1, 2, ..., N$, and we have taken the boundary conditions $\psi(z_0) = \psi(z_N) = 0$, since we expect the wavefunction to vanish at the boundaries z = 0 and z = $z_{max}$. To get the non-dimensional mass eigenvalues $\lambda_{\mu}$, we solve for the eigenvalue of the matrix $H_{\mu}$ given by
\begin{equation}
    H_{\mu} = \frac{1}{h^2}
    \begin{pmatrix}
        2 + h^2 V_{\mu}(z_1) & -1 & 0 & \cdots & 0 \\
        -1 & 2 + h^2 V_{\mu}(z_2) & -1 & \cdots & 0 \\
        0 & -1 & 2 + h^2 V_{\mu}(z_3) & \cdots & 0 \\
        \vdots & \vdots & \vdots & \ddots & \vdots \\
        0 & 0 & 0 & \cdots & 2 + h^2 V_{\mu}(z_{N-1}).
    \end{pmatrix}
\end{equation}
We use PyTorch's built-in function \texttt{torch.linalg.eigvalsh} to compute the eigenvalues of this real symmetric tridiagonal matrix, which returns the eigenvalues in ascending order. Under the hood, PyTorch utilizes the LAPACK library (or its GPU counterparts) to efficiently compute eigenvalues. Since the LAPACK algorithm used is not automatically differentiable, PyTorch's \texttt{torch.linalg.eigvalsh} uses a pre-programmed mathematical formula to compute the gradients of the eigenvalues with respect to the elements of the matrix. We derive a basic formula to calculate derivatives of eigenvalues of a real symmetric matrix \cite{PhysRevB.101.245139}.

Considering $A \in \mathbb{R}^{n \times n}$ to be a real symmetric matrix,
\begin{equation}
    \begin{split}
        A &= Q \Lambda Q^T, \\
        \Rightarrow AQ &= Q \Lambda,
    \end{split}
\end{equation}
where $Q \in \mathbb{R}^{n \times n}$ is an orthogonal matrix ($Q^T Q = I$) and $\Lambda = \text{diag}\left( \lambda_1, \lambda_2, ... , \lambda_n \right)$. Differentiating both sides and multiplying by $Q^T$ from the left, we get
\begin{equation*}
    Q^T (dA)Q + Q^T A(dQ) = Q^T (dQ)\Lambda + (d\Lambda).
\end{equation*}
We define $\Omega = Q^T (dQ)$, which is a skew-symmetric matrix (since $Q^T Q = I \Rightarrow Q^T (dQ) + (dQ)^T Q = 0 \Rightarrow \Omega^T = -\Omega$).
Substituting this in the above equation, we get
\begin{equation*}
    Q^T (dA)Q + \Lambda \Omega = \Omega \Lambda + d\Lambda,
\end{equation*} 
whose diagonal part is
\begin{equation*}
    d\lambda_i = q_i^T (dA) q_i,
\end{equation*}
where $q_i$ is the normalized eigenvector corresponding to the eigenvalue $\lambda_i$ (since each column of $Q$ is a normalized eigenvector of $A$). Using this, we can calculate the derivative of the eigenvalue with respect to any parameter $w$ as
\begin{equation}
    \frac{d\lambda_i}{dw} = q_i^T \left(\frac{dA}{dw}\right) q_i.
\end{equation}
This equation is the celebrated Hellmann-Feynman theorem \cite{PhysRev.56.340} in quantum mechanics that relates the derivative of an eigenvalue of a Hamiltonian to the expectation value of the derivative of the Hamiltonian operator.
PyTorch cleverly uses the eigenvectors $q_i$ to efficiently compute the gradients during the backward pass.

To get the mass eigenvalues of the pseudo-scalar meson $\pi$, we need to solve the coupled second-order differential equation given in Eq. \ref{eq:pi_mesons_equation}. The discretized form of the coupled equation is given by

\begin{equation}
    \begin{split}
        \frac{\pi_{i+1} - 2\pi_i + \pi_{i-1}}{h^2} + \omega(x_i) \frac{\pi_{i+1} - \pi_{i-1}}{2h} = m_n^2 (-\pi_i + \xi_i) \\
        C(z_i) \pi_i + \frac{\xi_{i+1} - 2\xi_i + \xi_{i-1}}{h^2} - B_1'(z_i) \frac{\xi_{i+1} - \xi_{i-1}}{2h} - C(z_i) \xi_i = 0,
    \end{split}
\end{equation}
where $B_1'(z) = \partial_z B_1(z)$, $\pi_i = \pi(z_i)$ and $\xi_i = \xi(z_i)$. 
From the Dirichlet boundary condition at $z = 0$, we have $\pi_0 = 0$ and $\xi_0 = 0$. From the Neumann boundary condition $\partial_z \pi = \partial_z \xi = 0$ at $z = z_{max}$, we have $\pi_N = \pi_{N-1}$ and $\xi_N = \xi_{N-1}$. Substituting these boundary conditions in the above equation, we get
\begin{equation}
    M \boldsymbol{\Psi} = m_n^2 T \boldsymbol{\Psi},
    \label{eq:generalised_eigenvalue_equation}
\end{equation}
where $\boldsymbol{\Psi} = (\pi_1, \pi_2, ..., \pi_{N-1}, \xi_1, \xi_2, ..., \xi_{N-1})^T$, and the matrices $M$ and $T$ can be written with block matrices as
\begin{equation}
    M = \begin{pmatrix}
        M_1 & 0 \\
        M_2 & M_3
    \end{pmatrix}, \quad
    T = \begin{pmatrix}
        -\mathds{I} & \mathds{I} \\
        0 & 0
    \end{pmatrix},
    \label{eq:matrix_M_T}
\end{equation}
where $\mathds{I}$ is the identity matrix of size $(N-1) \times (N-1)$, and the $M_1$, $M_2$ and $M_3$ are $(N-1) \times (N-1)$ block matrices given by,

    \begin{eqnarray}
        M_1 &= \begin{pmatrix}
            \frac{-2}{h^2} & \left(\frac{1}{h^2} + \frac{\omega(z_1)}{2h}\right) & 0 & \cdots & 0 \\
            \left(\frac{1}{h^2} - \frac{\omega(z_2)}{2h}\right) & \frac{-2}{h^2} & \left(\frac{1}{h^2} + \frac{\omega(z_2)}{2h}\right) & \cdots & 0 \\
            0 & \left(\frac{1}{h^2} - \frac{\omega(z_3)}{2h}\right) & \frac{-2}{h^2} & \cdots & 0 \\
            \vdots & \vdots & \vdots & \ddots & \vdots \\
            0 & 0 & 0 & \cdots & \frac{-2}{h^2} + \frac{1}{h^2} + \frac{\omega(z_{N-1})}{2h}
        \end{pmatrix}, 
    \end{eqnarray}
    \begin{eqnarray}
        M_2 &= \begin{pmatrix}
            C(z_1) & 0 & \cdots & 0 \\
            0 & C(z_2) & \cdots & 0 \\
            \vdots & \vdots & \ddots & \vdots \\
            0 & 0 & \cdots & C(z_{N-1})
            \end{pmatrix}, 
    \end{eqnarray}

    \begin{eqnarray}
        M_3 &= \begin{pmatrix}
            \frac{-2}{h^2} - C(z_1) & \frac{1}{h^2} - \frac{B_1'(z_1)}{2h} & 0 & \cdots & 0 \\
            \frac{1}{h^2} + \frac{B_1'(z_2)}{2h} & \frac{-2}{h^2} - C(z_2) & \frac{1}{h^2} - \frac{B_1'(z_2)}{2h} & \cdots & 0 \\
            0 & \frac{1}{h^2} + \frac{B_1'(z_3)}{2h} & \frac{-2}{h^2} - C(z_3) & \cdots & 0 \\
            \vdots & \vdots & \vdots & \ddots & \vdots \\
            0 & 0 & 0 & \cdots & \frac{-2}{h^2} - C(z_{N-1}) + \left(\frac{1}{h^2} + \frac{-B_1'(z_{N-1})}{2h}\right)
        \end{pmatrix}
    \end{eqnarray}

We solve the generalized eigenvalue equation given in Eq. \ref{eq:generalised_eigenvalue_equation} using \texttt{scipy.linalg.eigvals} giving us the non-dimensional mass eigenvalues $m_n^2$. The eigenvalues obtained are sorted in ascending order, and the first few eigenvalues are taken as the mass spectrum of the pseudo-scalar mesons. We have observed that, unlike the other mesons, the ground state of the $\pi$ meson needs a finer discretization to be accurately captured, which is why, for this case, we have chosen $h = 0.02$.

It is possible to convert the generalized eigenvalue equation in Eq. \ref{eq:generalised_eigenvalue_equation} to a standard eigenvalue equation by decoupling the equations
\begin{equation}
        M_1 \boldsymbol{\pi} = m_n^2 \mathds{I} (-\boldsymbol{\pi} + \boldsymbol{\xi}) \quad \text{and} \quad
        M_2 \boldsymbol{\pi} + M_3 \boldsymbol{\xi} = 0
\end{equation}
where $\boldsymbol{\pi} = (\pi_1, \pi_2, ..., \pi_{N-1})^T$ and $\boldsymbol{\xi} = (\xi_1, \xi_2, ..., \xi_{N-1})^T$. We can use the second equation to express either $\boldsymbol{\pi}$ or $\boldsymbol{\xi}$ in terms of the other variable, and substituting it in the first equation gives us a standard eigenvalue equation. However, this involves calculating the inverse of several matrices, which can introduce numerical errors. We have also observed that using this method leads to degeneracies in the eigenvalues, which are not observed when solving the generalized eigenvalue equation directly. Additionally, information about the eigenvectors is also lost in this process.
Hence, we choose to solve the generalized eigenvalue equation directly.

\section{Training details}
\label{appendix:training_details}
In this section, we provide the details of the training procedure and various hyperparameters used. We have taken $z_{max} = 15$ and $h = 0.15$, which gives $N = 100$ discretization points. The $a$ and $b$ values in Eq. \ref{eq:sigma_theta_rel} are taken as 0.01 GeV$^3$ and 0.03 GeV$^3$ respectively. 

For the neural networks used to approximate $A(z)$ and $v(z)$, we employed a fully connected feedforward network with four hidden layers, each containing 50 neurons. The input layer has one neuron (representing the variable $z$), and the output layer also has one neuron (representing either $A(z)$ or $v(z)$). The input is scaled by dividing by $z_{max}$ to ensure it lies within the range [0, 1].
We have used the \texttt{SiLU} (Swish) activation function \cite{ramachandran2017searchingactivationfunctions}, and the weights are initialized using the Kaiming normal initialization method \cite{he2015delvingdeeprectifierssurpassing} with leaky-ReLU. 

The $\Lambda_{\mu,n}$ corresponding to the weightage of each meson mass in the loss function, are adaptively chosen based on current loss values. 
The weightage is defined as,
\begin{equation}
    \Lambda_{\mu,n} = \frac{E_{\mu,n}}{\sum_{\mu \in}^{\{\rho, a_1, a_2, f_0\}} \sum_{i=1}^{N_{\mu}} E_{\mu,i}},
    \label{eq:Lambda_mass}
\end{equation}
where, $E_{\mu,n}$ is the error for the $n^{th}$ mass of the meson as given in Eq. \ref{eq:mass_error}. While calculating $\Lambda_{f_0,1}$, we scale the error $E_{f_0,1}$ by a factor of 0.1 to give it less weightage in the loss function, since the $f_0(500)$ meson is a broad resonance and its mass is not well defined.
Note that $\Lambda_{\mu,n}$ needs to be detached from the computational graph while calculating the loss to avoid any unwanted gradients.
Additionally we have chosen $\Lambda_{v'} = 10$, $\Lambda_{A} = 1$ and $\Lambda_{\text{pot}} = 1$  for the other loss components given in Eq. \ref{eq:loss_total}.

The initial value of the model parameters $l$ and $\theta$ is taken as $1.0$ and $0.0$, respectively. To find a good initial guess for the parameters $k_1$ and $k_2$, we employ a Bayesian optimization-based Tree-Structured Parzen Estimator (TPE) \cite{10.1145/3377930.3389817}. We use the \texttt{Optuna} library \cite{optuna_2019} to perform the optimization. The objective function for the optimization is defined as

\begin{equation}
    \mathcal{O}(k_1, k_2) = \sum_{\mu \in }^{\rho, a_1, a_2, f_0, \pi} \left(\frac{1}{N_{\mu}} \sum_{n=1}^{N_{\mu}} E_{\mu, n}\right).
\end{equation}

We define a search space for the initial value of $k_1$ and $k_2$ as $[-20, 20]$ and run the optimization for 100 trials. In each trial, we train the model for 10,000 iterations. The best initial guess values of $k_1$ and $k_2$ obtained from the optimization are $k_1 = -2.2$ and $k_2 = 12.0$. These values are used as the initial guess for $k_1$ and $k_2$ in the subsequent training runs.

To compute the loss $\mathcal{L}_{v'}$ given in Eq. \ref{eq:pos_v_dash_loss}, we choose 100 random points in the domain $z \in [0, z_{max}]$. For the loss $\mathcal{L}_{\text{pot}}$ given in Eq. \ref{eq:pos_pot_dash_loss}, we select one random point every iteration in the domain $z \in [0.8 z_{max}, z_{max}]$ to enforce the asymptotic behavior of the potential.

For the training, we have used two Adam optimizers \cite{kingma2017adammethodstochasticoptimization} with different learning rates. The first Adam optimizer is used to update the weights and biases of the neural network, with an initial learning rate of $1 \times 10^{-3}$, which is reduced in further iterations using a learning rate scheduler. The learning rate scheduler reduces the learning rate by a factor of 0.95 if the loss does not decrease for 100 iterations. The minimum learning rate is set to $10^{-8}$ for this optimizer.
The second Adam optimizer is used to train the model parameters $k_1$, $k_2$, $L$, and $\theta$, with an initial learning rate of $5\times10^{-2}$, along with a learning rate scheduler that reduces the learning rate by a factor of 0.6 if the loss does not decrease for 200 iterations. The minimum learning rate is set to $10^{-5}$ for this optimizer.

We have also implemented gradient clipping for all the parameters with a maximum norm of 1.0. Additionally, we have used double precision (float64) for better numerical stability. We train the model for 10,000 iterations.

\begin{figure}[!ht]
    \centering
    \includegraphics[width=0.8\textwidth]{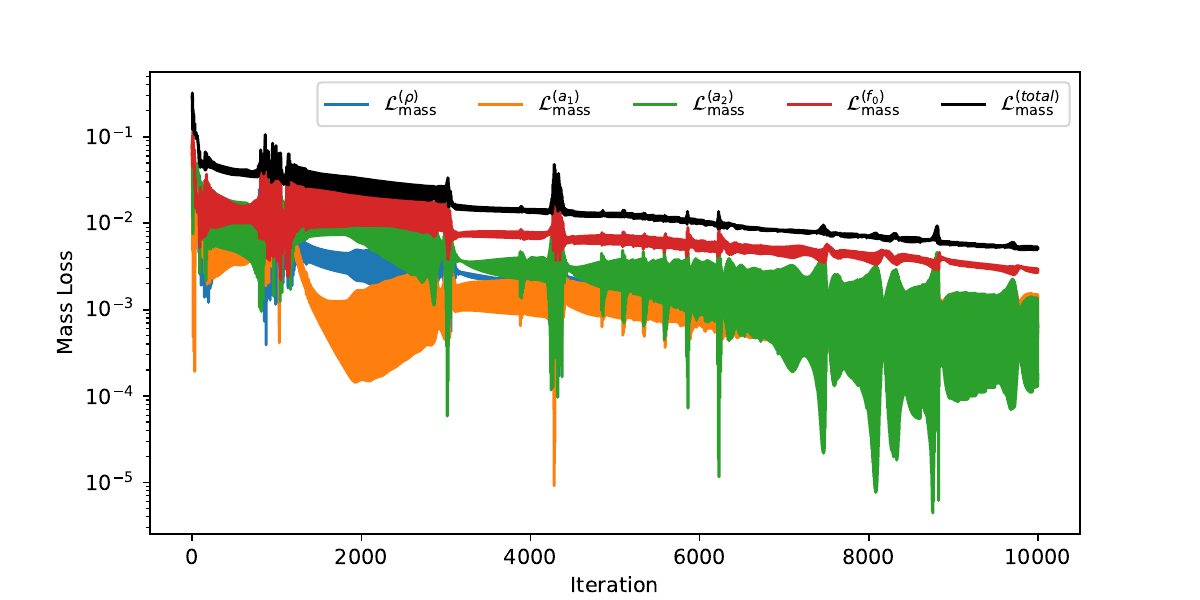}
    \caption{The loss curve for a representative training run.}
    \label{fig:loss_curve}
\end{figure}

\begin{table}[!ht]
\centering
\setlength{\tabcolsep}{6pt}
\renewcommand{\arraystretch}{1.2}
\begin{tabular}{|c|c|c|c|c|c|c|}
\hline
\textbf{Run} & min$(\mathcal{L}_{\text{mass}}^{(\rho, a_1, a_2, f_0)})$ & $\mathcal{L}_{\text{mass}}^{(\pi)}$ & $k_1$ & $k_2$ & $L$ (GeV)$^{-1}$ & $\theta$ \\
\hline
1 & $5.35\times 10^{-3}$ & 0.378 & $-4.01$ & $7.45$ & $1.23$ & $4.53$ \\
2 & $6.42\times 10^{-3}$ & 0.359 & $-4.84$ & $9.59$ & $1.20$ & $5.14$ \\
3 & $7.82\times 10^{-3}$ & 0.586 & $-3.51$ & $6.62$ & $1.60$ & $5.11$ \\
4 & $7.38\times 10^{-3}$ & 0.400 & $-2.93$ & $6.04$ & $1.00$ & $1.57$ \\
5 & $6.76\times 10^{-3}$ & 0.307 & $-2.54$ & $4.40$ & $1.00$ & $4.10$ \\
6 & $8.16\times 10^{-3}$ & 0.575 & $-4.15$ & $7.98$ & $1.61$ & $4.63$ \\
7 & $\mathbf{4.44\times 10^{-3}}$ & 0.510 & $-5.13$ & $10.16$ & $1.39$ & $4.56$ \\
8 & $4.94\times 10^{-3}$ & \textbf{0.305} & $-5.24$ & $10.75$ & $1.00$ & $3.33$ \\
9 & $6.89\times 10^{-3}$ & 0.474 & $-5.29$ & $10.87$ & $1.31$ & $2.12$ \\
10 & $4.72\times 10^{-3}$ & 0.353 & $-5.68$ & $12.13$ & $1.00$ & $1.88$ \\
\hline
\end{tabular}
\caption{Comparison of the minimum training loss for $\mathcal{L}_{\text{mass}}^{(\rho, a_1, a_2, f_0)}$ and the $\pi$ mass loss $\mathcal{L}_{\text{mass}}^{(\pi)}$ across all runs, along with the corresponding model parameters $(k_1, k_2, L, \theta)$. Run 7 achieved the lowest training loss $\mathcal{L}_{\text{mass}}^{(\rho, a_1, a_2, f_0)}$, while Run 8 yielded the lowest $\pi$ mass loss $\mathcal{L}_{\text{mass}}^{(\pi)}$, despite not being trained for it.}
\label{tab:training_loss}
\end{table}

We ran the training multiple times and chose only those runs where the minimum total loss converges to a value below $10^{-2}$. The loss curve for a representative run is shown in Fig. \ref{fig:loss_curve}. The training loss for each run is given in Table \ref{tab:training_loss}. We trained the model on an Intel Xeon W-2245 CPU (3.90 GHz, 16 GB RAM), which took around 20 minutes per run.

\bibliographystyle{JHEPCust}
\bibliography{ref}

\providecommand{\href}[2]{#2}\begingroup\raggedright\begin{thebibliography}{10}

\bibitem{Maldacena:1997re}
J.~M. Maldacena, \emph{{The Large $N$ limit of superconformal field theories
  and supergravity}},
  \href{http://dx.doi.org/10.4310/ATMP.1998.v2.n2.a1}{\emph{Adv. Theor. Math.
  Phys.} {\bf 2} (1998) 231--252},
  [\href{http://arxiv.org/abs/hep-th/9711200}{{\tt hep-th/9711200}}].

\bibitem{Gubser:1998bc}
S.~S. Gubser, I.~R. Klebanov and A.~M. Polyakov, \emph{{Gauge theory
  correlators from noncritical string theory}},
  \href{http://dx.doi.org/10.1016/S0370-2693(98)00377-3}{\emph{Phys. Lett. B}
  {\bf 428} (1998) 105--114}, [\href{http://arxiv.org/abs/hep-th/9802109}{{\tt
  hep-th/9802109}}].

\bibitem{Witten:1998qj}
E.~Witten, \emph{{Anti de Sitter space and holography}},
  \href{http://dx.doi.org/10.4310/ATMP.1998.v2.n2.a2}{\emph{Adv. Theor. Math.
  Phys.} {\bf 2} (1998) 253--291},
  [\href{http://arxiv.org/abs/hep-th/9802150}{{\tt hep-th/9802150}}].

\bibitem{Sakai:2004cn}
T.~Sakai and S.~Sugimoto, \emph{{Low energy hadron physics in holographic
  QCD}}, \href{http://dx.doi.org/10.1143/PTP.113.843}{\emph{Prog. Theor. Phys.}
  {\bf 113} (2005) 843--882}, [\href{http://arxiv.org/abs/hep-th/0412141}{{\tt
  hep-th/0412141}}].

\bibitem{Sakai:2005yt}
T.~Sakai and S.~Sugimoto, \emph{{More on a holographic dual of QCD}},
  \href{http://dx.doi.org/10.1143/PTP.114.1083}{\emph{Prog. Theor. Phys.} {\bf
  114} (2005) 1083--1118}, [\href{http://arxiv.org/abs/hep-th/0507073}{{\tt
  hep-th/0507073}}].

\bibitem{Kruczenski:2003uq}
M.~Kruczenski, D.~Mateos, R.~C. Myers and D.~J. Winters, \emph{{Towards a
  holographic dual of large N(c) QCD}},
  \href{http://dx.doi.org/10.1088/1126-6708/2004/05/041}{\emph{JHEP} {\bf 05}
  (2004) 041}, [\href{http://arxiv.org/abs/hep-th/0311270}{{\tt
  hep-th/0311270}}].

\bibitem{Hashimoto:2008sr}
K.~Hashimoto, T.~Hirayama, F.-L. Lin and H.-U. Yee, \emph{{Quark Mass
  Deformation of Holographic Massless QCD}},
  \href{http://dx.doi.org/10.1088/1126-6708/2008/07/089}{\emph{JHEP} {\bf 07}
  (2008) 089}, [\href{http://arxiv.org/abs/0803.4192}{{\tt 0803.4192}}].

\bibitem{Holdom:1982ex}
B.~Holdom and M.~E. Peskin, \emph{{Raising the Axion Mass}},
  \href{http://dx.doi.org/10.1016/0550-3213(82)90228-0}{\emph{Nucl. Phys. B}
  {\bf 208} (1982) 397--412}.

\bibitem{Holdom:1985vx}
B.~Holdom, \emph{{Strong QCD at High-energies and a Heavy Axion}},
  \href{http://dx.doi.org/10.1016/0370-2693(85)90371-5}{\emph{Phys. Lett. B}
  {\bf 154} (1985) 316}. [Erratum: Phys.Lett.B 156, 452 (1985)].

\bibitem{Erlich_2005}
J.~Erlich, E.~Katz, D.~T. Son and M.~A. Stephanov, \emph{Qcd and a holographic
  model of hadrons},
  \href{http://dx.doi.org/10.1103/physrevlett.95.261602}{\emph{Physical Review
  Letters} {\bf 95} (Dec., 2005) }.

\bibitem{Hirn:2005vk}
J.~Hirn, N.~Rius and V.~Sanz, \emph{{Geometric approach to condensates in
  holographic QCD}},
  \href{http://dx.doi.org/10.1103/PhysRevD.73.085005}{\emph{Phys. Rev. D} {\bf
  73} (2006) 085005}, [\href{http://arxiv.org/abs/hep-ph/0512240}{{\tt
  hep-ph/0512240}}].

\bibitem{Karch:2006pv}
A.~Karch, E.~Katz, D.~T. Son and M.~A. Stephanov, \emph{{Linear confinement and
  AdS/QCD}}, \href{http://dx.doi.org/10.1103/PhysRevD.74.015005}{\emph{Phys.
  Rev. D} {\bf 74} (2006) 015005},
  [\href{http://arxiv.org/abs/hep-ph/0602229}{{\tt hep-ph/0602229}}].

\bibitem{Csaki:2006ji}
C.~Csaki and M.~Reece, \emph{{Toward a systematic holographic QCD: A Braneless
  approach}},
  \href{http://dx.doi.org/10.1088/1126-6708/2007/05/062}{\emph{JHEP} {\bf 05}
  (2007) 062}, [\href{http://arxiv.org/abs/hep-ph/0608266}{{\tt
  hep-ph/0608266}}].

\bibitem{Falkowski:2006uy}
A.~Falkowski and M.~Perez-Victoria, \emph{{Holography, pade approximants and
  deconstruction}},
  \href{http://dx.doi.org/10.1088/1126-6708/2007/02/086}{\emph{JHEP} {\bf 02}
  (2007) 086}, [\href{http://arxiv.org/abs/hep-ph/0610326}{{\tt
  hep-ph/0610326}}].

\bibitem{Shock:2006gt}
J.~P. Shock, F.~Wu, Y.-L. Wu and Z.-F. Xie, \emph{{AdS/QCD Phenomenological
  Models from a Back-Reacted Geometry}},
  \href{http://dx.doi.org/10.1088/1126-6708/2007/03/064}{\emph{JHEP} {\bf 03}
  (2007) 064}, [\href{http://arxiv.org/abs/hep-ph/0611227}{{\tt
  hep-ph/0611227}}].

\bibitem{Karch:2010eg}
A.~Karch, E.~Katz, D.~T. Son and M.~A. Stephanov, \emph{{On the sign of the
  dilaton in the soft wall models}},
  \href{http://dx.doi.org/10.1007/JHEP04(2011)066}{\emph{JHEP} {\bf 04} (2011)
  066}, [\href{http://arxiv.org/abs/1012.4813}{{\tt 1012.4813}}].

\bibitem{Gherghetta_2009}
T.~Gherghetta, J.~I. Kapusta and T.~M. Kelley, \emph{Chiral symmetry breaking
  in the soft-wall ads/qcd model},
  \href{http://dx.doi.org/10.1103/physrevd.79.076003}{\emph{Physical Review D}
  {\bf 79} (Apr., 2009) }.

\bibitem{Zhang_2010}
P.~Zhang, \emph{Mesons and nucleons in soft-wall ads/qcd},
  \href{http://dx.doi.org/10.1103/physrevd.82.094013}{\emph{Physical Review D}
  {\bf 82} (Nov., 2010) }.

\bibitem{PhysRevD.81.014024}
Y.-Q. Sui, Y.-L. Wu, Z.-F. Xie and Y.-B. Yang, \emph{Prediction for the mass
  spectra of resonance mesons in the soft-wall ads/qcd model with a modified 5d
  metric}, \href{http://dx.doi.org/10.1103/PhysRevD.81.014024}{\emph{Phys. Rev.
  D} {\bf 81} (Jan, 2010) 014024}.

\bibitem{PhysRevB.101.245139}
H.~Xie, J.-G. Liu and L.~Wang, \emph{Automatic differentiation of dominant
  eigensolver and its applications in quantum physics},
  \href{http://dx.doi.org/10.1103/PhysRevB.101.245139}{\emph{Phys. Rev. B} {\bf
  101} (Jun, 2020) 245139}.

\bibitem{PhysRevD.79.075019}
W.~de~Paula, T.~Frederico, H.~Forkel and M.~Beyer, \emph{Dynamical holographic
  qcd with area-law confinement and linear regge trajectories},
  \href{http://dx.doi.org/10.1103/PhysRevD.79.075019}{\emph{Phys. Rev. D} {\bf
  79} (Apr, 2009) 075019}.

\bibitem{PhysRevD.108.106016}
A.~Ballon-Bayona, T.~Frederico, L.~A.~H. Mamani and W.~de~Paula,
  \emph{Dynamical holographic qcd model for spontaneous chiral symmetry
  breaking and confinement},
  \href{http://dx.doi.org/10.1103/PhysRevD.108.106016}{\emph{Phys. Rev. D} {\bf
  108} (Nov, 2023) 106016}.

\bibitem{PhysRevD.100.106009}
L.~A.~H. Mamani, \emph{Conformal symmetry breaking in holographic qcd},
  \href{http://dx.doi.org/10.1103/PhysRevD.100.106009}{\emph{Phys. Rev. D} {\bf
  100} (Nov, 2019) 106009}.

\bibitem{HORNIK1989359}
K.~Hornik, M.~Stinchcombe and H.~White, \emph{Multilayer feedforward networks
  are universal approximators},
  \href{http://dx.doi.org/https://doi.org/10.1016/0893-6080(89)90020-8}{\emph{Neural
  Networks} {\bf 2} (1989) 359--366}.

\bibitem{Hashimoto:2019bih}
K.~Hashimoto, \emph{{AdS/CFT correspondence as a deep Boltzmann machine}},
  \href{http://dx.doi.org/10.1103/PhysRevD.99.106017}{\emph{Phys. Rev. D} {\bf
  99} (2019) 106017}, [\href{http://arxiv.org/abs/1903.04951}{{\tt
  1903.04951}}].

\bibitem{Jeong:2025omu}
H.-S. Jeong, H.~Kim, K.-Y. Kim, G.~Yun, H.~Yu and K.~Yun,
  \emph{{AdS/Deep-Learning made easy II: neural network-based approaches to
  holography and inverse problems}},
  \href{http://arxiv.org/abs/2511.22522}{{\tt 2511.22522}}.

\bibitem{Hashimoto:2018bnb}
K.~Hashimoto, S.~Sugishita, A.~Tanaka and A.~Tomiya, \emph{{Deep Learning and
  Holographic QCD}},
  \href{http://dx.doi.org/10.1103/PhysRevD.98.106014}{\emph{Phys. Rev. D} {\bf
  98} (2018) 106014}, [\href{http://arxiv.org/abs/1809.10536}{{\tt
  1809.10536}}].

\bibitem{Akutagawa_2020}
T.~Akutagawa, K.~Hashimoto and T.~Sumimoto, \emph{Deep learning and ads/qcd},
  \href{http://dx.doi.org/10.1103/physrevd.102.026020}{\emph{Physical Review D}
  {\bf 102} (July, 2020) }.

\bibitem{Hashimoto_2021}
K.~Hashimoto, H.-Y. Hu and Y.-Z. You, \emph{Neural ordinary differential
  equation and holographic quantum chromodynamics},
  \href{http://dx.doi.org/10.1088/2632-2153/abe527}{\emph{Machine Learning:
  Science and Technology} {\bf 2} (may, 2021) 035011}.

\bibitem{Hashimoto:2021ihd}
K.~Hashimoto, K.~Ohashi and T.~Sumimoto, \emph{{Deriving the dilaton potential
  in improved holographic QCD from the meson spectrum}},
  \href{http://dx.doi.org/10.1103/PhysRevD.105.106008}{\emph{Phys. Rev. D} {\bf
  105} (2022) 106008}, [\href{http://arxiv.org/abs/2108.08091}{{\tt
  2108.08091}}].

\bibitem{PhysRevResearch.2.023369}
H.-Y. Hu, S.-H. Li, L.~Wang and Y.-Z. You, \emph{Machine learning holographic
  mapping by neural network renormalization group},
  \href{http://dx.doi.org/10.1103/PhysRevResearch.2.023369}{\emph{Phys. Rev.
  Res.} {\bf 2} (Jun, 2020) 023369}.

\bibitem{Filev:2025mbt}
V.~G. Filev, \emph{{Holographic flavour and neural networks}},
  \href{http://dx.doi.org/10.1007/JHEP11(2025)031}{\emph{JHEP} {\bf 11} (2025)
  031}, [\href{http://arxiv.org/abs/2506.20115}{{\tt 2506.20115}}].

\bibitem{Ahn:2024jkk}
B.~Ahn, H.-S. Jeong, K.-Y. Kim and K.~Yun, \emph{{Holographic reconstruction of
  black hole spacetime: machine learning and entanglement entropy}},
  \href{http://dx.doi.org/10.1007/JHEP01(2025)025}{\emph{JHEP} {\bf 01} (2025)
  025}, [\href{http://arxiv.org/abs/2406.07395}{{\tt 2406.07395}}].

\bibitem{wyr3-2kc5}
X.~Chen, Y.~Chen and K.~Zhou, \emph{Data-driven einstein-dilaton model for pure
  yang-mills thermodynamics and glueball spectrum},
  \href{http://dx.doi.org/10.1103/wyr3-2kc5}{\emph{Phys. Rev. D} {\bf 112}
  (Dec, 2025) 126025}.

\bibitem{Luo:2024iwf}
O.-Y. Luo, X.~Chen, F.-P. Li, X.-H. Li and K.~Zhou, \emph{{Neural network
  modeling of heavy-quark potential from holography}},
  \href{http://dx.doi.org/10.1140/epjc/s10052-025-14319-2}{\emph{Eur. Phys. J.
  C} {\bf 85} (2025) 637}, [\href{http://arxiv.org/abs/2408.03784}{{\tt
  2408.03784}}].

\bibitem{zhang2026heavyquarkoniumspectrumdecay}
Y.~Zhang, X.~Chen and M.~A.~M. Contreras, \emph{Heavy quarkonium spectrum and
  decay constants from a neural-network-based holographic model},  2026.

\bibitem{kou2026probingprotonstructurephysicsguided}
W.~Kou and X.~Chen, \emph{Probing proton structure via physics-guided neural
  networks in holographic qcd},  2026.

\bibitem{Cai:2024eqa}
R.-G. Cai, S.~He, L.~Li and H.-A. Zeng, \emph{{Neural ordinary differential
  equations for mapping the magnetic QCD phase diagram via holography}},
  \href{http://dx.doi.org/10.1007/s11433-025-2914-x}{\emph{Sci. China Phys.
  Mech. Astron.} {\bf 69} (2026) 240414},
  [\href{http://arxiv.org/abs/2406.12772}{{\tt 2406.12772}}].

\bibitem{ParticleDataGroup:2024cfk}
{\bf Particle Data Group} collaboration, S.~Navas et~al., \emph{{Review of
  particle physics}},
  \href{http://dx.doi.org/10.1103/PhysRevD.110.030001}{\emph{Phys. Rev. D} {\bf
  110} (2024) 030001}.

\bibitem{BaBar:2007ceh}
{\bf BaBar} collaboration, B.~Aubert et~al., \emph{{Measurements of $e^{+}
  e^{-} \to K^{+} K^{-} \eta$, $K^{+} K^{-} \pi^0$ and $K^0_{s} K^\pm \pi^\mp$
  cross- sections using initial state radiation events}},
  \href{http://dx.doi.org/10.1103/PhysRevD.77.092002}{\emph{Phys. Rev. D} {\bf
  77} (2008) 092002}, [\href{http://arxiv.org/abs/0710.4451}{{\tt 0710.4451}}].

\bibitem{SND:2023gan}
{\bf SND} collaboration, M.~N. Achasov et~al., \emph{{Study of the process $e^+
  e^- \to \omega \pi^0 \to \pi^+ \pi^- \pi^0 \pi^0$ in the energy range 1.05 --
  2.00 GeV with SND}},
  \href{http://dx.doi.org/10.1103/PhysRevD.108.092012}{\emph{Phys. Rev. D} {\bf
  108} (2023) 092012}, [\href{http://arxiv.org/abs/2309.00280}{{\tt
  2309.00280}}].

\bibitem{PhysRevLett.86.770}
{\bf Fermilab E791 Collaboration} collaboration, E.~M. Aitala, S.~Amato, J.~C.
  Anjos, J.~A. Appel, D.~Ashery, S.~Banerjee et~al., \emph{Experimental
  evidence for a light and broad scalar resonance in
  ${\mathit{d}}^{+}\ensuremath{\rightarrow}{\ensuremath{\pi}}^{\ensuremath{-}}{\ensuremath{\pi}}^{+}{\ensuremath{\pi}}^{+}$
  decay}, \href{http://dx.doi.org/10.1103/PhysRevLett.86.770}{\emph{Phys. Rev.
  Lett.} {\bf 86} (Jan, 2001) 770--774}.

\bibitem{WA76:1991kef}
{\bf WA76} collaboration, T.~A. Armstrong et~al., \emph{{Study of the centrally
  produced pi pi and K anti-K systems at 85-GeV/c and 300-GeV/c}},
  \href{http://dx.doi.org/10.1007/BF01548557}{\emph{Z. Phys. C} {\bf 51} (1991)
  351--364}.

\bibitem{Sarantsev:2021ein}
A.~V. Sarantsev, I.~Denisenko, U.~Thoma and E.~Klempt, \emph{{Scalar isoscalar
  mesons and the scalar glueball from radiative $J/\psi$ decays}},
  \href{http://dx.doi.org/10.1016/j.physletb.2021.136227}{\emph{Phys. Lett. B}
  {\bf 816} (2021) 136227}, [\href{http://arxiv.org/abs/2103.09680}{{\tt
  2103.09680}}].

\bibitem{BESIII:2022zel}
{\bf BESIII} collaboration, M.~Ablikim et~al., \emph{{Partial wave analysis of
  $J/\psi \rightarrow \gamma \eta' \eta'$}},
  \href{http://dx.doi.org/10.1103/PhysRevD.105.072002}{\emph{Phys. Rev. D} {\bf
  105} (2022) 072002}, [\href{http://arxiv.org/abs/2201.09710}{{\tt
  2201.09710}}].

\bibitem{RepoName}
R.~Pal, \emph{Github}, {\emph{https://github.com/rp-winter/NN-AdS-QCD} (2025)
  }.

\bibitem{Arkani-Hamed:2001kyx}
N.~Arkani-Hamed, A.~G. Cohen and H.~Georgi, \emph{{(De)constructing
  dimensions}},
  \href{http://dx.doi.org/10.1103/PhysRevLett.86.4757}{\emph{Phys. Rev. Lett.}
  {\bf 86} (2001) 4757--4761}, [\href{http://arxiv.org/abs/hep-th/0104005}{{\tt
  hep-th/0104005}}].

\bibitem{Cheng:2001nh}
H.-C. Cheng, C.~T. Hill and J.~Wang, \emph{{Dynamical Electroweak Breaking and
  Latticized Extra Dimensions}},
  \href{http://dx.doi.org/10.1103/PhysRevD.64.095003}{\emph{Phys. Rev. D} {\bf
  64} (2001) 095003}, [\href{http://arxiv.org/abs/hep-ph/0105323}{{\tt
  hep-ph/0105323}}].

\bibitem{Abe:2002rj}
H.~Abe, T.~Kobayashi, N.~Maru and K.~Yoshioka, \emph{{Field localization in
  warped gauge theories}},
  \href{http://dx.doi.org/10.1103/PhysRevD.67.045019}{\emph{Phys. Rev. D} {\bf
  67} (2003) 045019}, [\href{http://arxiv.org/abs/hep-ph/0205344}{{\tt
  hep-ph/0205344}}].

\bibitem{Randall:2002qr}
L.~Randall, Y.~Shadmi and N.~Weiner, \emph{{Deconstruction and gauge theories
  in AdS(5)}},
  \href{http://dx.doi.org/10.1088/1126-6708/2003/01/055}{\emph{JHEP} {\bf 01}
  (2003) 055}, [\href{http://arxiv.org/abs/hep-th/0208120}{{\tt
  hep-th/0208120}}].

\bibitem{Falkowski:2002cm}
A.~Falkowski and H.~D. Kim, \emph{{Running of gauge couplings in AdS(5) via
  deconstruction}},
  \href{http://dx.doi.org/10.1088/1126-6708/2002/08/052}{\emph{JHEP} {\bf 08}
  (2002) 052}, [\href{http://arxiv.org/abs/hep-ph/0208058}{{\tt
  hep-ph/0208058}}].

\bibitem{Son:2003et}
D.~T. Son and M.~A. Stephanov, \emph{{QCD and dimensional deconstruction}},
  \href{http://dx.doi.org/10.1103/PhysRevD.69.065020}{\emph{Phys. Rev. D} {\bf
  69} (2004) 065020}, [\href{http://arxiv.org/abs/hep-ph/0304182}{{\tt
  hep-ph/0304182}}].

\bibitem{deBlas:2006fz}
J.~de~Blas, A.~Falkowski, M.~Perez-Victoria and S.~Pokorski, \emph{{Tools for
  deconstructing gauge theories in AdS(5)}},
  \href{http://dx.doi.org/10.1088/1126-6708/2006/08/061}{\emph{JHEP} {\bf 08}
  (2006) 061}, [\href{http://arxiv.org/abs/hep-th/0605150}{{\tt
  hep-th/0605150}}].

\bibitem{Erlich:2006hq}
J.~Erlich, G.~D. Kribs and I.~Low, \emph{{Emerging holography}},
  \href{http://dx.doi.org/10.1103/PhysRevD.73.096001}{\emph{Phys. Rev. D} {\bf
  73} (2006) 096001}, [\href{http://arxiv.org/abs/hep-th/0602110}{{\tt
  hep-th/0602110}}].

\bibitem{Nakai:2014iea}
Y.~Nakai, \emph{{Deconstruction, Holography and Emergent Supersymmetry}},
  \href{http://dx.doi.org/10.1007/JHEP03(2015)101}{\emph{JHEP} {\bf 03} (2015)
  101}, [\href{http://arxiv.org/abs/1412.3486}{{\tt 1412.3486}}].

\bibitem{KIRITSIS200767}
E.~Kiritsis and F.~Nitti, \emph{On massless 4d gravitons from asymptotically
  ads5 space–times},
  \href{http://dx.doi.org/https://doi.org/10.1016/j.nuclphysb.2007.02.024}{\emph{Nuclear
  Physics B} {\bf 772} (2007) 67--102}.

\bibitem{Chelabi_2016}
K.~Chelabi, Z.~Fang, M.~Huang, D.~Li and Y.-L. Wu, \emph{Chiral phase
  transition in the soft-wall model of ads/qcd},
  \href{http://dx.doi.org/10.1007/jhep04(2016)036}{\emph{Journal of High Energy
  Physics} {\bf 2016} (Apr., 2016) 1–30}.

\bibitem{U_Gursoy_2008}
U.~Gürsoy and E.~Kiritsis, \emph{Exploring improved holographic theories for
  qcd: part i},
  \href{http://dx.doi.org/10.1088/1126-6708/2008/02/032}{\emph{Journal of High
  Energy Physics} {\bf 2008} (feb, 2008) 032}.

\bibitem{PhysRevC.79.045203}
A.~Cherman, T.~D. Cohen and E.~S. Werbos, \emph{Chiral condensate in
  holographic models of qcd},
  \href{http://dx.doi.org/10.1103/PhysRevC.79.045203}{\emph{Phys. Rev. C} {\bf
  79} (Apr, 2009) 045203}.

\bibitem{PhysRevD.83.016002}
T.~M. Kelley, S.~P. Bartz and J.~I. Kapusta, \emph{Pseudoscalar mass spectrum
  in a soft-wall model of ads/qcd},
  \href{http://dx.doi.org/10.1103/PhysRevD.83.016002}{\emph{Phys. Rev. D} {\bf
  83} (Jan, 2011) 016002}.

\bibitem{PhysRevD.102.026013}
A.~Ballon-Bayona and L.~A.~H. Mamani, \emph{Nonlinear realization of chiral
  symmetry breaking in holographic soft wall models},
  \href{http://dx.doi.org/10.1103/PhysRevD.102.026013}{\emph{Phys. Rev. D} {\bf
  102} (Jul, 2020) 026013}.

\bibitem{10.1145/3377930.3389817}
Y.~Ozaki, Y.~Tanigaki, S.~Watanabe and M.~Onishi, \emph{Multiobjective
  tree-structured parzen estimator for computationally expensive optimization
  problems},  in \emph{Proceedings of the 2020 Genetic and Evolutionary
  Computation Conference}, GECCO '20, (New York, NY, USA), p.~533–541,
  Association for Computing Machinery, 2020.
\newblock \href{http://dx.doi.org/10.1145/3377930.3389817}{DOI}.

\bibitem{optuna_2019}
T.~Akiba, S.~Sano, T.~Yanase, T.~Ohta and M.~Koyama, \emph{Optuna: A
  next-generation hyperparameter optimization framework},  in \emph{Proceedings
  of the 25th {ACM} {SIGKDD} International Conference on Knowledge Discovery
  and Data Mining}, 2019.

\bibitem{PhysRevD.98.046019}
K.~Hashimoto, S.~Sugishita, A.~Tanaka and A.~Tomiya, \emph{Deep learning and
  the $\mathrm{AdS}/\mathrm{CFT}$ correspondence},
  \href{http://dx.doi.org/10.1103/PhysRevD.98.046019}{\emph{Phys. Rev. D} {\bf
  98} (Aug, 2018) 046019}.

\bibitem{PhysRev.56.340}
R.~P. Feynman, \emph{Forces in molecules},
  \href{http://dx.doi.org/10.1103/PhysRev.56.340}{\emph{Phys. Rev.} {\bf 56}
  (Aug, 1939) 340--343}.

\bibitem{ramachandran2017searchingactivationfunctions}
P.~Ramachandran, B.~Zoph and Q.~V. Le, \emph{Searching for activation
  functions},  2017.

\bibitem{he2015delvingdeeprectifierssurpassing}
K.~He, X.~Zhang, S.~Ren and J.~Sun, \emph{Delving deep into rectifiers:
  Surpassing human-level performance on imagenet classification},  2015.

\bibitem{kingma2017adammethodstochasticoptimization}
D.~P. Kingma and J.~Ba, \emph{Adam: A method for stochastic optimization},
  2017.

\end{thebibliography}\endgroup

\end{document}